\documentclass[a4paper,11pt]{article}
\usepackage{jheppub}
\usepackage{amsmath,amssymb,slashed,makecell}

\author[a]{Martina Ferrillo}
\affiliation[a]{Physik-Institut, Universität Zürich, Zürich, Switzerland}
\author[b,c]{Maksym Ovchynnikov}
\affiliation[b]{Institut für Astroteilchen Physik, Karlsruher Institut für Technologie (KIT), Hermann-von-Helmholtz-Platz 1, 76344 Eggenstein-Leopoldshafen, Germany
}
\affiliation[c]{Instituut-Lorentz for Theoretical Physics, Universiteit Leiden, Niels Bohrweg 2, 2333 CA Leiden, The Netherlands}
\author[d]{Filippo Resnati}
\author[d]{Albert De Roeck}
\affiliation[d]{CERN, Geneva, Switzerland}

\emailAdd{martina.ferrillo@cern.ch}
\emailAdd{maksym.ovchynnikov@cern.ch}
\emailAdd{albert.de.roeck@cern.ch}
\emailAdd{Filippo.Resnati@cern.ch}
\begin{document}

\title{Improving the potential of BDF@SPS to search for new physics with liquid argon time projection chambers}

\abstract
{Beam dump experiments proposed at the SPS are perfectly suited to explore the parameter space of models with long-lived particles, thanks to the combination of a large intensity with a high proton beam energy. In this paper, we study how the exploration power may be augmented further by installing a detector based on liquid argon time projection chamber technology. In particular, we consider several signatures of new physics particles that may be uniquely searched for with such a detector, including double bang events with heavy neutral leptons, inelastic light dark matter, and millicharged particles.}

\maketitle
\flushbottom

\section{Introduction}
\label{sec:introduction}

Feebly-interacting particles, or FIPs, are hypothetical particles with a mass below the electroweak scale and couplings to SM particles that are sufficiently small to be unconstrained by previous experiments. Depending on the FIPs' properties, they may resolve present problems in the Standard Model, such as neutrino oscillations, dark matter, and the baryon asymmetry of the Universe. 

The interest in FIPs has increased significantly 
over the last decade~\cite{Beacham:2019nyx,Antel:2023hkf}, resulting in various experiments being proposed to search for them. Assuming the FIP mass range is  $\mathcal{O}(1 -10\text{ GeV})$, a perfect facility for such experiments is the CERN SPS, since it delivers a proton beam of relatively high energy of $E_{p} = 400\text{ GeV}$ with a huge proton intensity. In collisions with a target, FIPs may be copiously produced and detected in downstream experiments.

Three experiments have recently been proposed to be installed at the ECN3 facility at SPS: SHiP~\cite{Aberle:2839677}, SHADOWS~\cite{Alviggi:2839484}, and HIKE~\cite{CortinaGil:2839661} (see also the recent report~\cite{Ahdida:2867743}). At the time of this writing, the selection and reviewing process of these proposals is ongoing. HIKE may operate in two modes: the kaon mode, which would explore new physics emerging in rare processes with kaons, and the beam dump mode, which would allow the search for decays of long-lived FIPs. SHiP and SHADOWS, equipped with a hidden sector decay spectrometer and -- in the case of SHiP -- a scattering and neutrino detector (SND), could probe the FIPs by their decay and scattering processes.

In this paper, we argue that the FIP exploration to be delivered with the described setups does not fully use the potential of the facility. We show that it may be significantly extended if installing an additional liquid argon (LAr) detector based on the time projection chamber technology (LArTPC). Thanks to the timing capabilities, low recoil threshold, and fully electronic equipment, it would complement the abilities of the decay spectrometer and the SND and allow the search for FIPs by utilizing unique signatures that are inaccessible with the mentioned detectors. In this study, we will consider SHiP as the experiment to host the LAr detector, although, in principle, any of the proposed experiments may be equipped with it if there is available space.\footnote{Currently, there are LArTPC prototype detectors already installed at the SPS -- the so-called ProtoDUNE detector~\cite{DUNE:2020cqd}. The potential of ProtoDUNE for searches for FIPs is discussed in~\cite{Coloma:2023adi}.}

The paper is organized as follows. In Sec.~\ref{sec:experiments}, we briefly describe the SHiP setup, overview the SND@SHiP detector, and discuss a possible extension with a LArTPC setup in detail. In Sec.~\ref{sec:LAr-opportunities}, we discuss the new opportunities that may be delivered with LAr@SHiP. Secs.~\ref{sec:ldm},~\ref{sec:mcp},~\ref{sec:ldm-inelastic},~\ref{sec:dipole} are devoted to the discussion of the physics reach of LAr@SHiP for particular models with FIPs. Finally, in Sec.~\ref{sec:conclusions}, we make conclusions.

\section{The SHiP experiment and LArTPC detector}
\label{sec:experiments}

\begin{figure}
    \centering
    \includegraphics[width=0.98\textwidth]{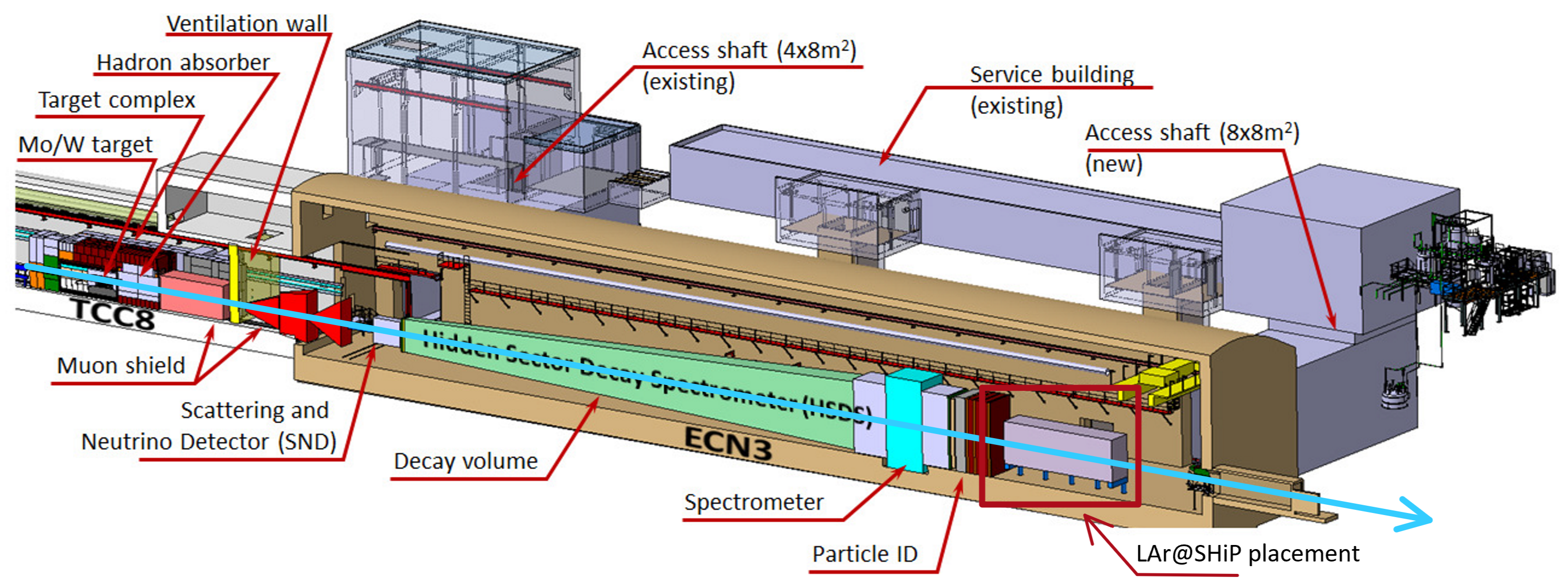}
    \caption{The layout of the SHiP experiment. The figure is taken from~\cite{Aberle:2839677}. The direction of the incoming proton beam is indicated by the cyan line, whereas the possible placement of the LAr@SHiP is shown by the red box (with the LAr volume depicted in pink).}
    \label{fig:SHIPdetector}
\end{figure}

SHiP~\cite{Ahdida:2704147, Aberle:2839677, Ahdida:2654870, Ship_det_paper} is a beam dump experiment proposed to be installed at the ECN3 facility at SPS, see Fig.~\ref{fig:SHIPdetector}. It combines the detector setup, which is close to optimal in maximization of the new physics particle event rate~\cite{Bondarenko:2023fex}, with the suppression of backgrounds down to a negligible level. It consists of the target made of tungsten and molybdenum, the hadron absorber followed by the magnetic deflector (called the muon shield), the scattering and neutrino detector SND@SHiP, the 50 meters long hidden sector decay volume, and the 15-meter long hidden sector decay products detector that includes the particle identification systems. SND@SHiP would study neutrino physics and search for the scattering of new physics particles, while the decay volume would look for their decay. 

\subsection{SND@SHiP}
\begin{figure}
    \centering
    \includegraphics[width=0.95\textwidth]{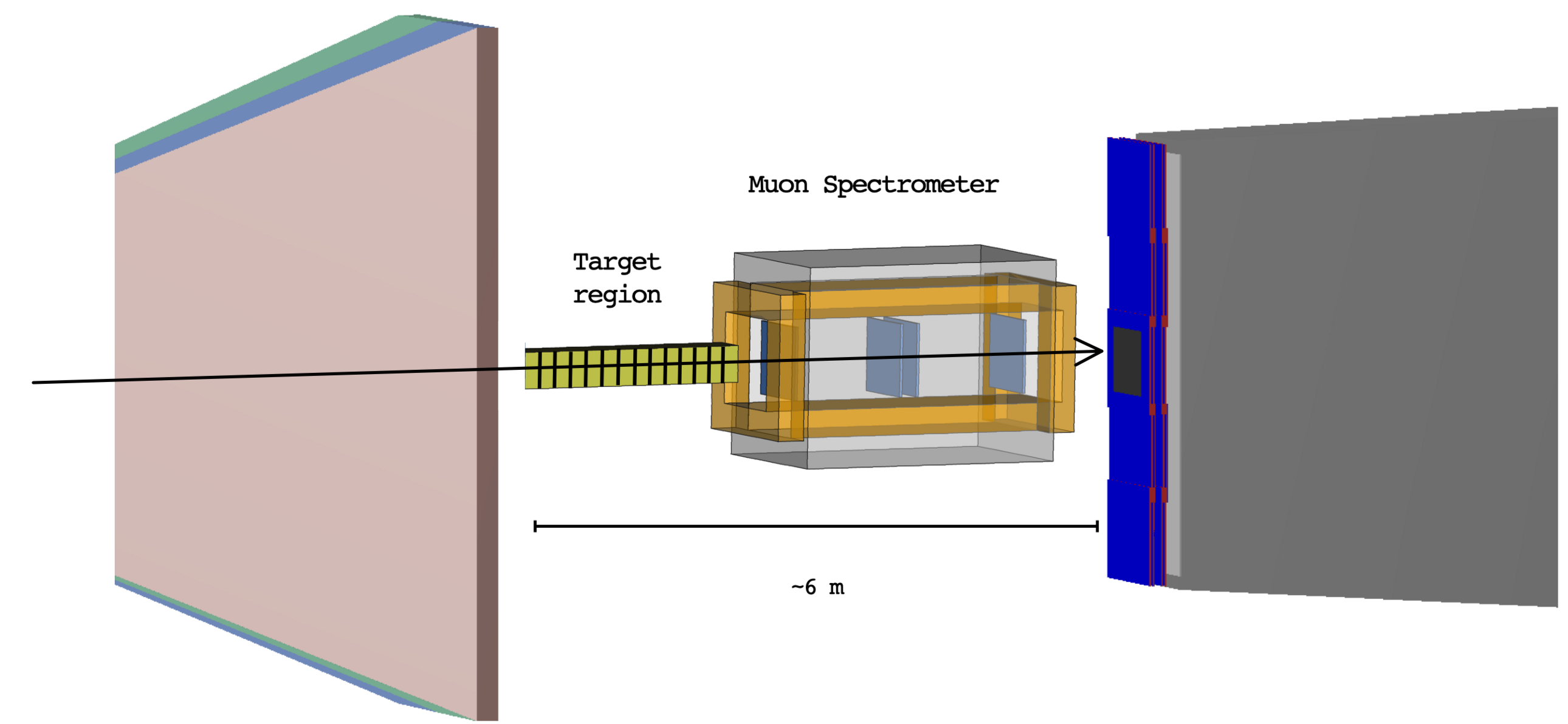}
    \caption{Conceptual layout of the Scattering and Neutrino Detector at SHiP (SND@SHiP) in the ECN3 configuration. It is located between the muon shield (the pink object on the left) and the hidden sector decay spectrometer (the gray object on the right). The upstream part of SND@SHiP is the neutrino/FIPs target region and vertex detector, while the downstream one is the muon spectrometer. The black line with the arrow indicates the beam direction.}
    \label{fig:SNDdetector}
\end{figure}

The Scattering and Neutrino Detector (SND@SHiP) was specifically designed to identify interactions of neutrinos of all flavours and scattering of Feebly Interacting Particles (FIPs) such as light dark matter, originating from the proton beam dump and subsequent interactions~\cite{Ahdida:2867743,  SHiP_ECN3}. Its modular layout, schematically shown in Fig.~\ref{fig:SNDdetector} as implemented in the Monte Carlo simulation of the experiment, includes a combined neutrino/LDM target and vertex detector upstream based on the Emulsion Cloud Chamber (ECC) technology~\cite{Acquafredda_2009}, followed by a Muon Spectrometer for the measurement of the charge and momentum of muons produced in $\nu_{\mu}$ Charged Current (CC) interactions and in the muonic decay channel of the tau produced in $\nu_{\tau}$ CC interactions. What follows is a concise description of the detector's main features corresponding to the baseline configuration extensively detailed in Refs.~\cite{Ahdida:2704147, Aberle:2839677, Ahdida:2654870, Ship_det_paper, SHiP_ECN3}, which was used for the studies presented in this work.\\
\indent The ECC section of the SND@SHiP is composed of an alternation of tungsten layers as passive absorber and nuclear emulsion films acting as high granularity tracking devices, resulting in a detector with sub-micrometric position and milli-radian angular resolution as shown by the OPERA~\cite{ref:Acquafredda} and SND@LHC~\cite{SNDLHC:2022ihg} experiments. Each elementary ECC unit, a brick, consists of 60 nuclear emulsion films of $20\times20\,\mathrm{cm^2}$ cross-sectional area, interleaved with 59 tungsten plates with a thickness of $1\,\mathrm{mm}$, corresponding to a total weight of $\sim 45\,\mathrm{kg}$ and $\sim7.8\,\mathrm{cm}$ thickness. Given the sub-micrometric spatial resolution in an ECC brick, the momentum measurement of charged particles is possible via the detection of their multiple Coulomb scattering in the absorber~\cite{Agafonova_2012}. In addition, each ECC brick acts as a high granularity sampling calorimeter with $\sim 1\,X_{0}$ every three sensitive layers. ECC bricks are assembled in 17 walls made by $2\times2$ ECC units each for a total length of $2.6\,\mathrm{m}$ and fiducial mass of $\sim 3$ tonnes.\\
\indent The ECC target walls are alternated with electronic detector tracking planes, the Target Trackers (TT), with the main task of locating the position of the interaction happening within the emulsion target while complementing the electromagnetic showers energy measurement. Furthermore, TT particle tracks can be linked with those reconstructed in the emulsions and in the muon spectrometer, helping with the identification of muons from $\nu_{\mu}$ interactions and muonic decays of the $\tau$ lepton.  With its $100\,\mu$m position and $\sim250\,$ps time resolution, the Scintillating Fibre (SciFi) tracker technology, already in use in the SND@LHC experiment~\cite{SNDLHC:2022ihg}, represents a valid option under consideration for the TT detector.\\
\indent A Muon Spectrometer, equipped with four tracking stations situated in a $1\,\mathrm{T}$ magnetic field, is located downstream of the neutrino/FIPs target area. Its role is to measure the momentum of muons in combination with the Hidden Sector Decay Spectrometer (HSDS), placed immediately downstream of the SND@SHiP detector.\\
\indent As a result, the SND@SHiP detector is ideally suited to reconstruct interaction vertices of neutrinos of all flavours and topologically disentangle them from the decay of short-lived particles (\textit{e.g.} $\tau$ leptons, charged hadrons)~\cite{PhysRevLett.115.121802, ref:opera} and scattering vertices of FIPs off the nucleons and electrons of the passive material.\\

\subsection{LAr@SHiP}
\label{sec:lar}
An interesting detector technology to complement and enhance the capabilities for searches for new physics particles is that of a Liquid Argon Time Projection Chamber (LArTPC). LArTPCs are imaging and homogeneous calorimetric devices that are very suitable as detectors for rare event searches. The LArTPCs output is digitized bubble-chamber-like images that can be tridimensionally reconstructed, allowing to distinguish between different interaction processes with high accuracy. Photodetectors recording the scintillation light are typically used for triggering the detector and fast timing information.

This LArTPC technology has matured a lot over the last ten years and is now regularly used as a technology for neutrino detectors and dark matter search experiments. Most notably, the ICARUS LArTPC of about 500 tons was originally one of the far detectors at the LNGS for the CNGS neutrino beam~\cite{ICARUS:2023gpo}. ICARUS is now exploited as the far detector in the short baseline neutrino oscillation experiment at Fermilab, together with SBND~\cite{SBND:2020scp}, another LArTPC, as a near detector. Four 10 kTon active mass LArTPCs will be used as far detectors for the DUNE experiment~\cite{DUNE:2020lwj}. Also, the Forward Physics Facility, a proposal being prepared for forward physics studies at the LHC, plans to include a large LArTPC experiment called FLArE~\cite{Feng:2022inv}. At CERN, there is significant experience with building the large 700-ton LArTPC detectors that are constructed as prototypes for the large DUNE far detectors~\cite{DUNE:2020cqd}.

LArTPCs provide an actual electronic event picture of the signal candidates of interest that decay or scatter in their fiducial volume. E.g., for a
Heavy Neutral Lepton (HNL) decaying in the detector, the decay vertex and tracks and/or showers coming from the decay particles can be reconstructed (we will return to this in Sec.~\ref{sec:LAr-opportunities}). Similarly, e.g., light dark matter particles or millicharged particles (MCPs) produced in the beam dump target that scatter with the argon atoms of the detector lead to visible signals.

Recently, LArTPCs have been used for searches for millicharged~\cite{ArgoNeuT:2019ckq} particles, heavy QCD axions~\cite{ArgoNeuT:2022mrm}, HNLs~\cite{ArgoNeuT:2021clc,MicroBooNE:2022ctm,MicroBooNE:2023icy} and Higgs portal scalars~\cite{MicroBooNE:2022ctm} in ArgoNeuT and MicroBooNE. MeV-scale energy depositions by low-energy photons produced in neutrino-argon interactions have been identified and reconstructed in ArgoNeuT liquid argon time projection chamber data. Analyses are presently ongoing in ICARUS on (light) Dark Matter searches, and have been reported by dedicated Dark Matter experiments such as DarkSide~\cite{DarkSide:2022knj}. Future neutrino experiments such as SBND (starting in 2024) and the DUNE experiment, in particular via the near detector, will have LArTPCs to address new physics searches.

For SHIP, a possible configuration is to install a LArTPC behind the spectrometer, where a $\approx 23$ long free space will be available; see Fig.~\ref{fig:SHIPdetector}. Such a detector will enhance the SHiP physics program with sensitivity to light dark matter scattering and millicharged particles passing the detector, as well as complement the searches for decays of HNLs, axions, dark photons, and more. No version of a LArSHiP detector has been included in the simulation yet, so these studies represent initial results. Clearly, if an excess is observed in any of these channels in the experiment, a visual confirmation of the observation in, e.g., a LArTPC will be of paramount importance to strengthen the case for discovery.


The critical TPC components are 1) the HV system, in charge of creating a stable and uniform electric field throughout the active volume, 2) the charge readout modules, for which several technologies and geometry (wire, strips, pixels, \dots) exist and have been tested in multiple detectors, 3) the photon detector system to record the scintillation light signals, 4) sensitive and low noise electronics for preamplification of the charge signals, and 5) the data acquisition and triggering system.

For what concerns the infrastructure, the LArTPC requires the cryostat that contains the detector components and the liquid argon and limits the heat input and the cryogenics system in charge of maintaining stable thermodynamic conditions and achieving sufficient argon purity.

\begin{figure}[!t]
\centering
\includegraphics[width=0.70\linewidth]{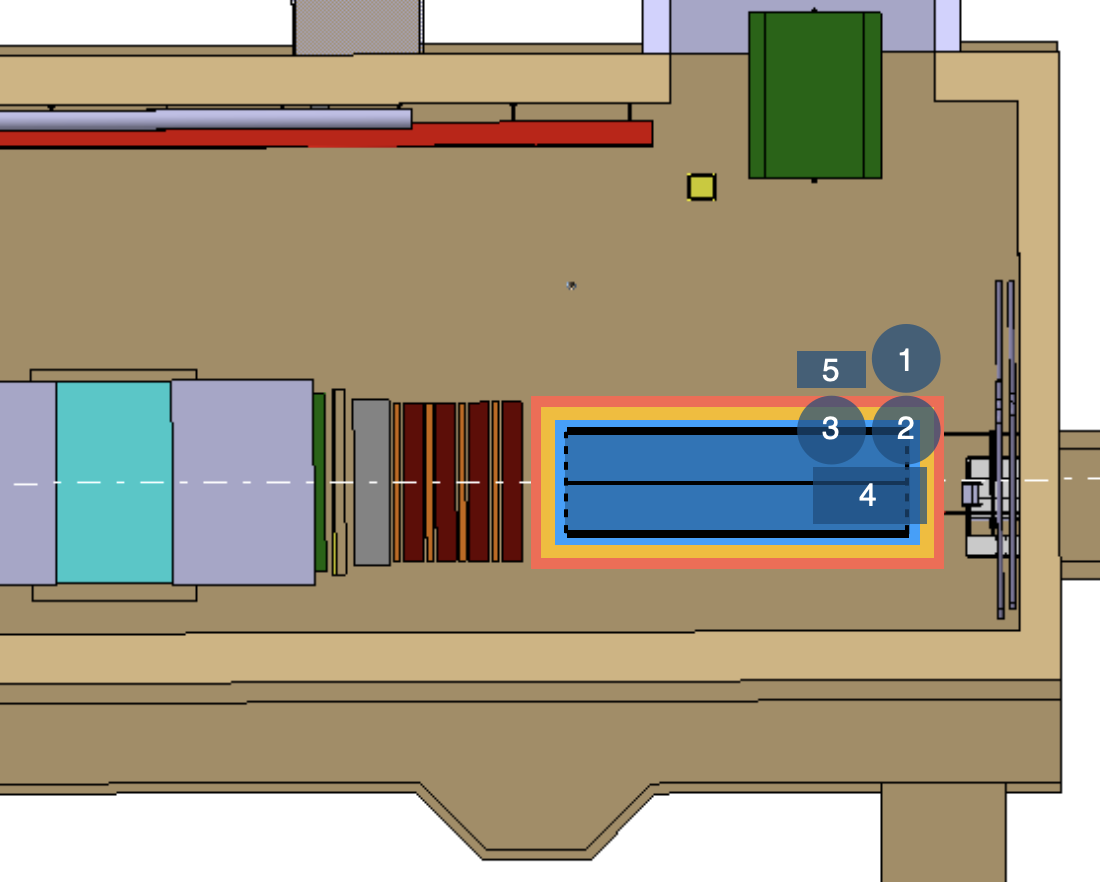}
\caption{
Top view of the space needed for a 3x3x10~m$^3$ LAr TPC and related infrastructure. The cryostat is represented in red and yellow, and the liquid argon is represented in blue. Two TPC volumes are shown. Space for the proximity cryogenic is also shown: 1) LAr pump box, 2) LAr condenser box, 3) phase separator, 4) filtration plant, and 5) warm gas management system. Equipment in 2), 3), and 4) should be installed above the cryostat.
}
\label{Fig:LArTPC}
\end{figure}

For SHiP, an LArTPC based on the following configuration could be envisaged. The space available behind SHiP has a footprint that allows the installation of a TPC with an active volume up to $3 \times 3 \times 10$ m$^3$ (about 130 tons) and its cryogenic system. The volume could be split into two TPCs, each one with a drift length of 1.5 m and a drift time of approximately 1 ms. Such a layout is shown in Fig.~\ref{Fig:LArTPC}. Further details, such as the granularity of the readout volume (e.g., one large volume or divided into cells), need to be studied and optimized with detailed simulations, and which will respond to possible issues of pile-up from background cosmic ray muons and muons from the beam dump that evade the upstream magnetic shield. 

In Table~\ref{tab:snd-vs-lar}, we summarize the geometric parameters of SND@SHiP and LAr@SHiP, where for SND, we also include the old setup considered for the ECN4 cavern, for which the sensitivity to light dark matter has been calculated in detail~\cite{SHiP:2020noy}.

\begin{table}[t!]
    \centering
    \begin{tabular}{|c|c|c|c|c|c|c|c|c|c|}
     \hline  Setup  &  $\frac{z_{\text{to det}}}{\text{m}}$  &  Target material& $\frac{m_{\text{det}}}{t}$ & $\frac{\Delta z_{\text{tg}}}{\text{m}}$ & $\frac{(\Delta x \times \Delta y)_{\text{tg}}}{\text{m}^{2}}$ &  $\frac{\Omega}{\text{sr}}$    \\ 
     \hline
      SND@SHiP@ECN4  &   38  &  Lead & 8   &  $1$   &  $0.9\times 0.75$  &  $4.7\cdot 10^{-4}$     \\
     \hline
      SND@SHiP@ECN3  &  25   & Tungsten   & 3 &  $1$  & $0.4\times 0.4$   & $2.6\cdot 10^{-4}$      \\ \hline
       LAr@SHiP &  97   & LAr  & 130 &  $10$  &  $3\times 3$  &  $9.6\cdot 10^{-4}$      \\ \hline
    \end{tabular}
    \caption{Parameters of the setups of the scatterings detectors considered in this paper: the old SND@SHiP@ECN4 configuration used in~\cite{SHiP:2020noy} to calculate the SHiP sensitivity to LDM, its updated setup for SHiP@ECN3 described in~\cite{Aberle:2839677}, and LAr@SHiP. The meaning of the parameters is as follows: the longitudinal distance from the target to the beginning of the detector, the detector material, the target mass, the longitudinal dimension of the target, its transverse dimensions, and the solid angle covered by the detector.}
    \label{tab:snd-vs-lar}
\end{table}

\section{New signatures to be explored with LAr}
\label{sec:LAr-opportunities}

LAr@SHiP may provide opportunities complementary to the abilities of the HSDS and SND to explore FIP decay and scattering signatures, as well as exploit signatures that would be very challenging with the latter detectors.

For decays of FIPs, the event sensitivity of LAr@SHiP could exceed one of the HSDS due to the geometric limitations of the latter, while for the LArTPC, the decay products are observed at the decay vertex in the LArTPC. However, the detector is placed farther downstream, has a smaller angular coverage, and its effective decay volume length is smaller. However, thanks to a very precise spatial resolution and the decay volume being a fully electronic read-out detector, it has important advantages in event reconstruction. The trade-off of these advantages and disadvantages will need to be studied in detail in future work.

\begin{figure}[t!]
    \centering
    \includegraphics[width=0.65\textwidth]{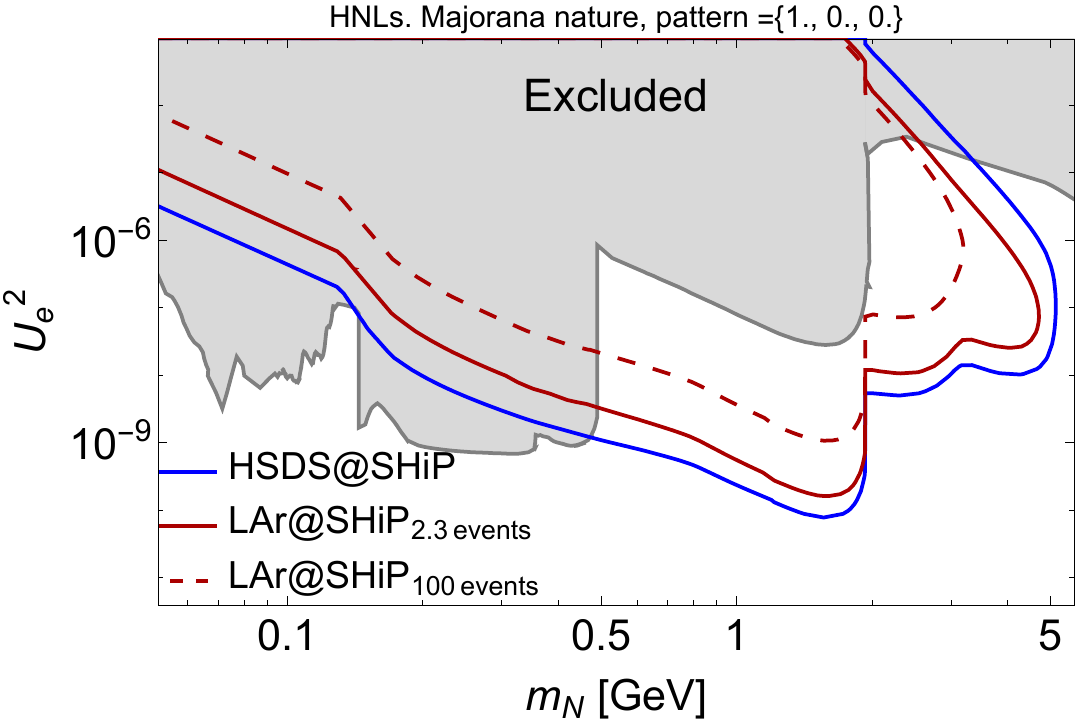}
    \caption{The parameter space of the HSDS@SHiP sensitivity where the FIP decay events may be visualized at LAr@SHiP assuming $N_{\text{PoT}} = 6\cdot 10^{20}$ (which is equivalent to 15 years of SHiP running), using HNLs coupled to electron neutrinos as an example. The blue line shows the 90\% CL sensitivity of HSDS@SHiP, while the red lines are the iso-contours indicating the domain where LAr may be able to see 2.3 (solid) and 100 (dashed) events, assuming unit reconstruction efficiency. The curves have been obtained using \texttt{SensCalc}~\cite{Ovchynnikov:2023cry}. For the events selection at HSDS and LAr@SHiP, we followed the LoI~\cite{Aberle:2839677} and~\cite{Breitbach:2021gvv} correspondingly, see also Sec.~\ref{sec:selection-background}.}
    \label{fig:event-visualization-coverage}
\end{figure}

First, a LArTPC may serve as an event display, visualizing the FIP decay vertices~\cite{Adams:2020uco,DUNE:2022wlc}. This is especially important for decays of heavy FIPs $m_{\text{FIP}}\gtrsim 1\text{ GeV}$, which would typically have high multiplicities. It is complicated to fully reconstruct such events at 
the HSDS since many of the decay particles would escape the spectrometer coverage. The coverage of the parameter space covered by the  HSDS where LAr@SHiP would be able to visualize events is quite significant, as shown in Fig.~\ref{fig:event-visualization-coverage}.

Second, with a LAr detector, it may be possible to search FIPs by mono-particle decays, where only one of the decay particles or scattered particles is visible. An example is a decay into a photon and a neutrino. Searching for this decay at the HSDS would not be possible since one needs a pair of particles reaching the spectrometer to reconstruct the vertex position. This is not the case for the LAr detector since it will allow the reconstruction of the decay event directly at the decay vertex. We will discuss practical applications of such signatures in Sec.~\ref{sec:dipole} by considering the dipole portal of HNLs.

When considering FIPs scattering signatures, the use of a LArTPC detector might nicely complement the ECC technology of the SND@SHiP. The LAr detector features a lower detection energy threshold for recoil electrons, of the order of tens of MeV, against the $1\,\mathrm{GeV}$ needed to reconstruct electron-initiated electromagnetic showers within a single ECC brick in order to discard any integrated background.  In addition, the LAr technology has intrinsic time reconstruction capabilities, which are unavailable within the ECC itself but provided by Target Trackers in the SND@SHiP. As a consequence, the LAr setup is ideally suited for the detection of FIPs, whose scatterings proceed via the EM interaction or hypothetical interactions with light mediators $m\ll~1~\,\text{GeV}$, resulting in a final state low energy recoil electron. An advantage of integrated timing information resides in the opportunity to reconstruct sequential FIPs scattering signatures. We will consider the opportunities provided by low energy recoils for simple single scatterings of FIPs in Sec.~\ref{sec:ldm}. 

\begin{figure}[t!]
    \centering
    \includegraphics[width=0.5\textwidth]{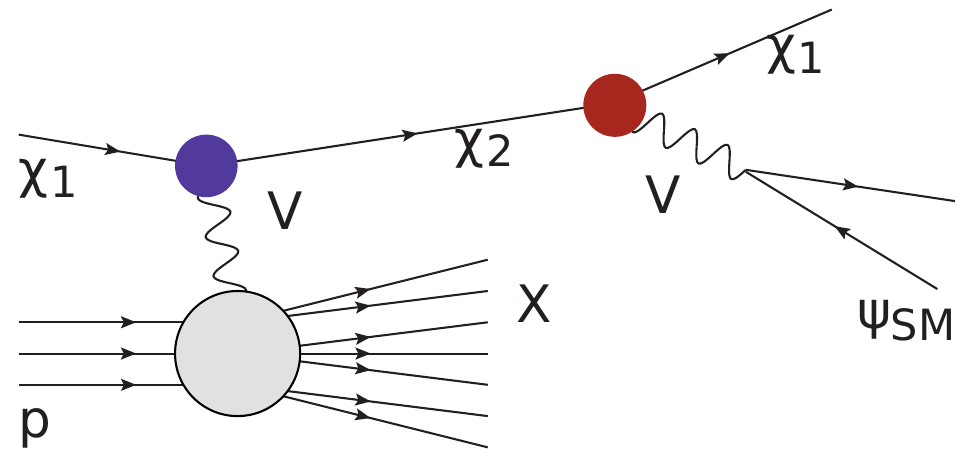}~\includegraphics[width=0.5\textwidth]{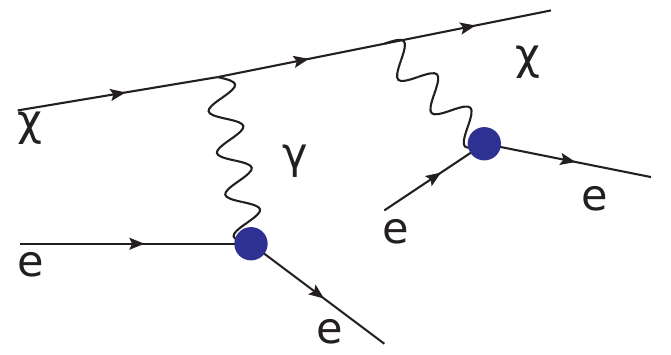}
    \caption{Examples of the double bang signatures that may be searched for at LAr@SHiP. \textit{Left panel}: A stable particle $\chi_{1}$ scatters off electrons or nucleons, producing a heavier unstable particle $\chi_{2}$ and a recoil SM particle. If $\chi_{2}$ is short-lived enough, the decay probability does not suppress the rate of such events. Being time-correlated, these events may be distinguished from backgrounds. \textit{Right panel}: if a stable FIP elastically scatters off SM particles via a light or massless mediator (such as the EM field), the recoil energy would be very low. Its smallness may compensate for the smallness of the FIP interaction coupling, and FIP may experience several low-energy recoil scatterings.}
    \label{fig:double-bang}
\end{figure}

Excellent timing and recoil threshold properties of LArTPC provide opportunities to use combined signatures inaccessible with HSDS and SND. An example of such signatures is the ``multi bang'' (DB) event, see Fig.~\ref{fig:double-bang}, where the origin of the bangs are various processes with FIPs inside the detector. We will concentrate on two different cases. In the first case (the right panel of the figure), a FIP would elastically scatter off SM particles with a light mediator. Low energy scattering recoils may parametrically compensate for the smallness of the interaction coupling, and the FIP may even have a chance to scatter several times before leaving the detector. Due to a low energy recoil, the line obtained by joining the scattering ``bangs'' is approximately straight and points to the FIP production point, which heavily simplifies the background rejection. We discuss this signature using the example of millicharged particles in Sec.~\ref{sec:mcp}.

The second example (the left panel of Fig.~\ref{fig:double-bang}) is when some stable FIP inelastically scatters off SM particles and produces another FIP, which then decays within the detector (which causes the second bang). Unlike the MCP DB signature, the line joining the two bangs would not closely point to the target since, in order to produce a particle with a different mass, one needs to generate a transverse momentum relative to the direction of motion of the incoming particle. Hence, the background rejection is more complicated. Fortunately, since the produced unstable particle has an energy well above a GeV, its decay products are energetic, so there is no need for tight energy thresholds for the second bang, which helps dealing with backgrounds. We will consider examples of the double bang events in Secs.~\ref{sec:ldm-inelastic},~\ref{sec:dipole} using the models of inelastic light dark matter and a dipole portal of HNLs.

The attractiveness of the DB signature is that it would give us much more information than single-bang events. For instance, when collecting a significant amount of events, we can determine the decay length and lifetime of the unstable particle, reconstruct its main decay modes, and thus, in this way, explore the properties of the particle. In addition, the signature would give us access to search for processes with relatively short lifetimes of the order of $c\tau \sim 1\text{ cm}$, which are otherwise inaccessible at beam dump experiments due to the long nominal distance from the dump to the detector.

\subsection{Backgrounds discussion}
\label{sec:selection-background}

There are two main sources of background at LAr@SHiP: interactions induced by cosmic rays and the SM particles produced in the dump and reaching the detector, mostly muons.

For the cosmic muons, using the estimates from~\cite{Coloma:2023adi} made for ProtoDUNE, taking into account that LAr@SHiP has the volume $\simeq 2.5$ times smaller and also the fact that SPS operates only $\sim 200$ days/year, we may estimate the number of muons that may cause these backgrounds as $4.4\cdot 10^{5}/\text{year}$, or $6.6\cdot 10^{6}$ per 15 years. This background may be significantly reduced using time synchronization with a beam-target collision; indeed, the proton-target collisions are split into spills, and the amount of spills is $10^{6}/\text{year}$. Further background reduction may be achieved by reconstructing the event. For instance, if the cosmic muon leaves a track image within the detector, one may reject the event by using the angular cut - a requirement on the direction made by the two hits to approximately point to the direction of the target. Even if the angular cut would not work, the event may be rejected using the hypothesis of a particular topology of events with new physics particles (which typically differs from the cosmic rays interactions topology).

\begin{figure}[t!]
    \centering
    \includegraphics[width=0.6\textwidth]{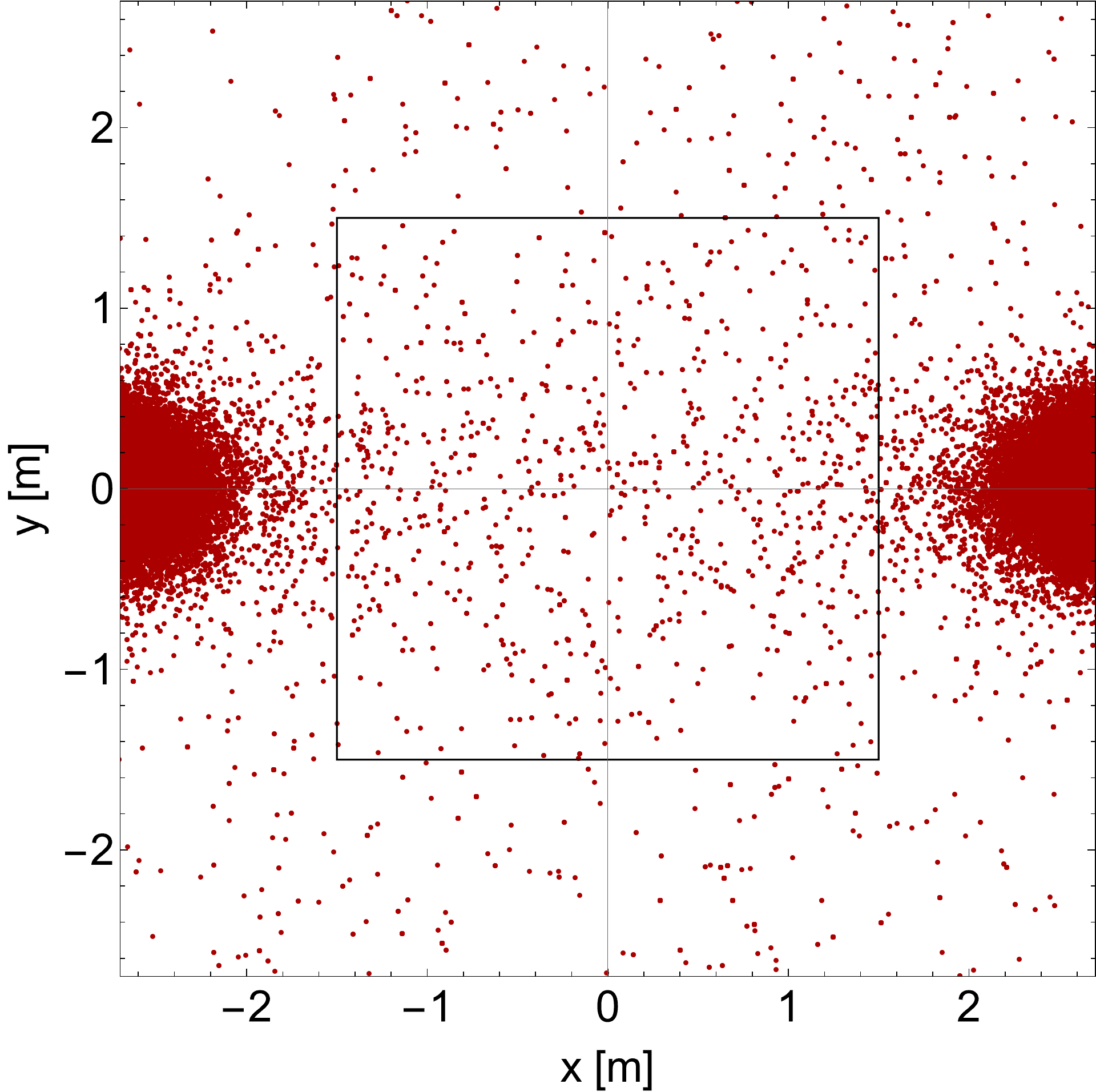}
    \caption{The distribution of muon hits per spill in the plane of the beginning of LAr-SHiP as obtained in SHiP simulations~\cite{Aberle:2839677}. The black box in the middle of the plot shows the boundary of LAr@SHiP. The effect of the muon shield causes the spots at the left-right edge of the plot.}
    \label{fig:muons-from-dump}
\end{figure}

For the beam dump muons, the situation is more complicated. Given the muon shield configuration for the current SHiP setup in~\cite{Aberle:2839677}, the number of muons crossing the LAr plane is $3.7\cdot 10^{3}$ per spill, see Fig.~\ref{fig:muons-from-dump}. The optimization of the muon shield is ongoing. Hence, the current rate is pessimistic and may well be significantly reduced, possibly by an order of magnitude. This background will, however, be significantly reduced using the upstream part of the LAr volume as a veto, as muons are detected as charged particles and leave track-like signatures. Then, similarly to the cosmic background, it may be further reduced using the angular cut and event reconstruction. Below, we optimistically assume zero background from muons, taking the first meter of the detector as a veto. The background resulting from muons interacting with material close by the LArTPC leading to neutral hadrons entering the active volume of the detector needs to be evaluated with detailed simulations, but their interactions would lead to unexpected z-dependence for a true signal. We also expect to control this background with the information of the HSDS in front of the LArTPC, where the tracks of the incoming muons would be registered before they interact with the material can be measured.\footnote{The same background rejection logic may be applied to the case of placements of LAr at SHADOWS and HIKE experiments, assuming that the LAr detector would be located downward their spectrometers.}

Another important background comes from neutrino scatterings. It is relevant mostly for mono-particle scattering signatures with a single electron or nucleon. We may roughly estimate the number of neutrino scatterings during a 15-year running time by knowing the number of neutrino interactions in SND@SHiP. Both the SND and LAr setups cover the far-forward angular region where the solid angle distribution of neutrinos is isotropic, and energies are similar. For the signal, we consider the recoil energy window $30\text{ MeV}<E_{e}<1\text{ GeV}$, based on the ability of the LAr detector to reconstruct low-energy events. At SND, counting neutrino interactions resulting in such an electron, we get $N_{\text{bg,SND}} \approx 312$ events (see Sec.~\ref{sec:ldm} for details on the simulation). At the LArTPC, we expect 
\begin{equation}
    N_{\text{bg,LAr}} \simeq N_{\text{bg,SND}}\times \frac{\Omega_{\text{LAr}}}{\Omega_{\text{SND}}}\times \frac{Z_{\text{tg}}^{\text{LAr}}}{Z_{\text{tg}}^{\text{SND}}}\cdot \frac{\Delta z_{\text{tg}}^{\text{LAr}}}{\Delta z_{\text{tg}}^{\text{SND}}} \approx 9N_{\text{bg,SND}} \approx 2.8\cdot 10^{3},
\end{equation}
where $\Omega$ scaling comes from the geometric acceptance, and $Z\cdot \Delta z$ from the scattering probability.

Deep inelastic neutrino scatterings may act as a background for decays of FIPs. However, they typically have a different topology -- in particular, the presence of nucleons among the recoil particles, (often) a higher multiplicity, and wider angular distribution of the decay products. The LArTPc may accurately reconstruct the events and use this difference to discriminate signals from the background.

\section{Light dark matter coupled to dark photons}
\label{sec:ldm}

The interaction sector in the model of the LDM $\chi$ coupled to dark photons $V$ is described by the Lagrangian
\begin{equation}
    \mathcal{L} = -\frac{\epsilon}{2}F_{\mu\nu}V^{\mu\nu} + |D_{\mu}\chi|^{2},
\end{equation}
where $D_{\mu}=\partial_{\mu}-ig_{D}V_{\mu}$ is the covariant derivative, $\epsilon$ is the mixing between the dark photon and the SM photon, and $g_{D}$ is the coupling of $\chi$ to $V$.

\begin{figure}[t!]
    \centering
    \includegraphics[width=0.65\textwidth]{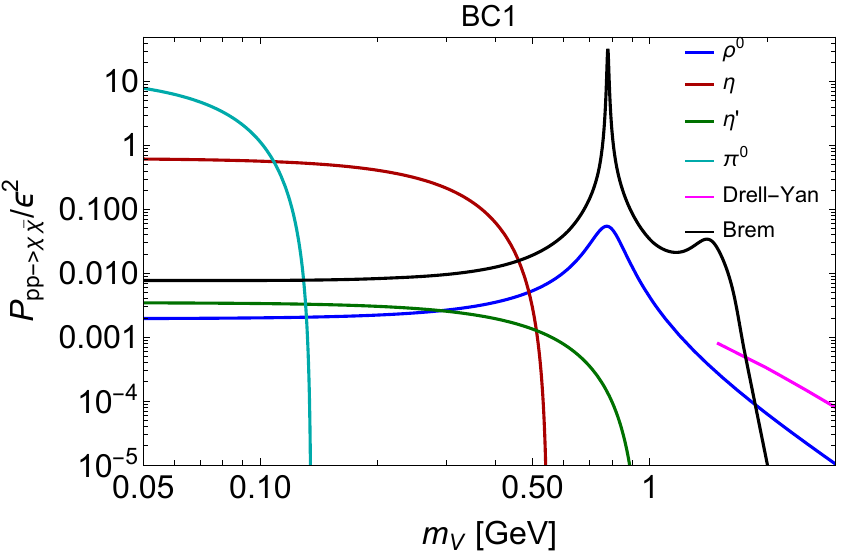}
    \caption{The production probability of the pair $\chi\bar{\chi}$ as the function of the dark photon mass $m_{V}$ if assuming that $\text{Br}(V\to \chi\bar{\chi})\approx 1$ at SPS. The channel are: the mixing of dark photons with $\rho^{0}$, decays of light mesons $\pi^{0},\eta,\eta'$, bremsstrahlung process, and the Drell-Yan process. For details, see Ref.~\cite{Ovchynnikov:2023cry}.}
    \label{fig:LDM-prod-probabilities}
\end{figure}

We will consider the mass range $m_{\chi}<m_{V}/2$, and a large $g_{D}\gg \epsilon$. In this case, the $\chi$ particles may be copiously produced by decays of dark photons, with $\text{Br}(V\to \chi\bar{\chi}) = 1$. The dark photon particles, in their turn, may be produced by deep inelastic scattering, proton bremsstrahlung, and decays of light mesons $\pi^{0},\eta,\eta'$. The total yields of the $\chi\bar{\chi}$ pairs produced in the collisions of the proton beam with the molybdenum/tungsten target are shown in Fig.~\ref{fig:LDM-prod-probabilities}.

The detection signature may be $\chi$s scattering off electrons, nuclei, nucleons, and deep-inelastic scattering. In this work, we concentrate on the elastic scattering off electrons, keeping in mind that the omitted channels may contribute significantly to the sensitivity. This way, our estimates are conservative. 

\subsection{SND@SHiP}
\label{sec:ldm-snd}
The main background to LDM scattering searches in SND@SHiP is dominated by neutrino interactions sharing the same event topology at the primary vertex. For the present work, we concentrate on LDM elastic scattering signatures off electrons inside the SND detector as in Ref.~\cite{SHiP:2020noy} for the ECN4 configuration. In this scenario, the signal features a single outgoing charged track being an electron. Abundantly produced in the beam dump upstream of the SND, neutrinos of all flavours with a single electron in the final state arise from several elastic and inelastic scattering processes with potentially multiple unreconstructed tracks. The relevant background processes are all summarized in Table.~\ref{tab:nuyield}.\\
The framework of the SHiP experiment~\cite{fairship} was used to produce Monte Carlo simulations for the neutrino background. Proton on target collisions were simulated by means of \texttt{Pythia~v8.23}~\cite{Sjostrand:2014zea}, while the detector geometry and transport via \texttt{GEANT4}~\cite{GEANT4:2002zbu}. Finally, the neutrino scatterings within the SND detector were produced with the \texttt{GENIE~v2.12.6} software~\cite{Andreopoulos:2009rq}.\\
We adopt a two-step approach to estimate the neutrino background for LDM elastic scattering searches, closely following Ref.~\cite{SHiP:2020noy}. At first, only neutrino interactions within the detector acceptance and with one visible track at the primary vertex are kept in the selection. A visibility criterion is applied to charged tracks to be reconstructed in the emulsion medium, corresponding to a momentum of 170 MeV/c for protons and 100 MeV/c for other charged particles~\cite{ref:operavis}. In addition, the presence of photons or $\pi^0$ near the interaction vertex is vetoed, further reducing the residual background.\\
The second step consists of a kinematic selection in the phase space of the scattered electron, energy $E_e$, and polar angle $\theta_e$, as it offers discriminatory power between the kinematics of neutrinos from LDM candidates. The optimal selection region is defined by maximizing the significance of the observation:
\begin{equation}
\label{eq:significance}
    \Sigma = \frac{S}{\sqrt{\sigma_{\rm stat}^2+\sigma_{\rm sys}^2}} = \frac{S}{\sqrt{ B + \sum\limits_{\substack{i \ell }} \left(\kappa_{i\ell} B_{i\ell}\right)^2}}\,,
\end{equation}
where the LDM signal is denoted by $S$, the total neutrino background by $B$. The systematic uncertainty on the neutrino fluxes and cross sections is hereby taken into account by using factors $\kappa_{i\ell} = \sqrt{\kappa_{i}
^2+\tilde{\kappa}_\ell^2}$, with the index $i$ summed over the neutrino flavor and $\ell$ over the neutrino interaction type. The relevant contributions to the systematic budget from the neutrino cross section are reported in Table~\ref{tab:xsecsys}. We assume the systematic uncertainty on the neutrino flux to be dominated by the precision on the neutrino Deep Inelastic scattering cross section at the level of $5\%$~\cite{Wu:2007ab, SHiP:2020noy}.\\
The optimized selection is identified via a grid-like scan of the significance $\Sigma$ in the kinematic region ($E_e$, $\theta_e$), yielding the phase space region $E_e\in[1, 5]\,$GeV and $\theta_e\in[10, 30]\,$mrad.\\
The estimate of the neutrino interactions, after the selection and corresponding to $N_{p.o.t.}=6\times 10^{20}$, is reported in Table~\ref{tab:nuyield}. We note that the residual background is represented by the irreducible elastic scatterings and quasi-elastic processes $\bar{\nu}_e\,p\to e^+\,n$.
\begin{table}[t]
\centering
\begin{tabular}{lccccc}
\hline
& $\nu_{e}$ & $\bar{\nu}_{e}$ & $\nu_{\mu}$ & $\bar{\nu}_{\mu}$ & all\\ 
\hline
Elastic scattering on $e^{-}$ & 260 & 135 & 320 & 210 & 925 \\
Quasi~-~elastic scattering  & - & 45 &  &  & 45 \\
Resonant scattering  & - & - &  &  & - \\
Deep inelastic scattering  & - & - &  &  & -\\
\hline
Total & 260 & 180 & 320 & 210 & 970 \\
\hline
\end{tabular}
\caption{Neutrino background yield corresponding to the full SHiP running time of 15 years, which is equivalent to $6\times 10^{20}$ delivered PoT, assuming the selection for LDM-electron elastic scattering searches.}
\label{tab:nuyield}
\end{table}

\begin{table}[t!]
    \centering
    \begin{tabular}{c  c}
         \hline
         Neutrino interaction & Systematic uncertainty\\
         \hline
         \hline
         Elastic scattering on $e^{-}$ & Negligible~\cite{RevModPhys.84.1307} \\
         Quasi~-~elastic scattering & $8\%$~\cite{Lyubushkin:2008pe}\\
         Resonant scattering & $18\%$~\cite{PhysRevD.83.052007}\\
         Deep inelastic scattering & $5\%$~\cite{Wu:2007ab}\\
         \hline
    \end{tabular}
    \caption{Systematic uncertainty on the neutrino cross section for relevant background processes to LDM elastic scattering searches.}
    \label{tab:xsecsys}
\end{table}

The sensitivity of SND@SHiP for the old ECN4 configuration has been calculated in Ref.~\cite{SHiP:2020noy} assuming $m_{\chi}/m_{V} = 1/3$ and the signature of scattering off electrons. Given the similarities between this past setup and the new SHiP setup to be operated at the ECN3 facility (see Table~\ref{tab:snd-vs-lar}), keeping unchanged the signature and the mass ratio, and knowing the background yields at these two setups --- 230~\cite{SHiP:2020noy} for the old ECN4 setup and 582 for the ECN3 setup (assuming $3$ times larger number of protons on target), the sensitivity of the new configuration may be obtained with the help of a simple rescaling. Namely, at the lower bound of the sensitivity, we get
\begin{equation}
    Y_{\text{lower}} \propto \left(\frac{A_{\text{tg}}\sqrt{N_{\text{bg}}}z^{2}_{\text{to det}}}{Z_{\text{tg}}m_{\text{det}}}\right)^{\frac{1}{2}},
    \label{eq:rescaling-1}
\end{equation}
where $Y \equiv \epsilon^{2}\alpha_{D}(m_{\chi}/m_{V})^{4}$ and $A_{\text{tg}}$ is the mass number of the target's nuclei. The details about the derivation of this formula are given in Appendix~\ref{app:rescaling}.

Plugging the numbers from Table~\ref{tab:snd-vs-lar} in this equation, we find that
\begin{equation}
    Y_{\text{lower,ECN3}} \approx Y_{\text{lower,ECN4}}\times 3^{-\frac{1}{4}}
\end{equation}
The sensitivity is shown in Fig.~\ref{fig:sensitivity-LDM}.

\subsection{LAr@SHiP}

Similarly to the SND@SHiP, the LAr setup is located in the far-forward direction. Therefore, it could be again possible to obtain its sensitivity using a rescaling of the SND@SHiP sensitivity. However, here we are interested in a completely different kinematic regime for LAr -- low energy recoil electrons with $30\text{ MeV}<E_{e}<1\text{ GeV}$ instead of $1\text{ GeV}<E_{e}<5\text{ GeV}$ (plus the angular cut) for the SND setup. Therefore, in~\eqref{eq:rescaling-1}, we have to include the additional factor $\sqrt{\sigma_{\text{SND}}/\sigma_{\text{LAr}}}$, where $\sigma_{\text{exp}}$ is the integrated cross-section for the phase space satisfying the selection for the given experiment. For the background, we will assume neutrino scattering only, with the total amount given by $N_{\text{bg}}\approx 2.8\cdot 10^{3}$ (sec.~\ref{sec:selection-background}).

\begin{figure}[t!]
    \centering
    \includegraphics[width=0.65\textwidth]{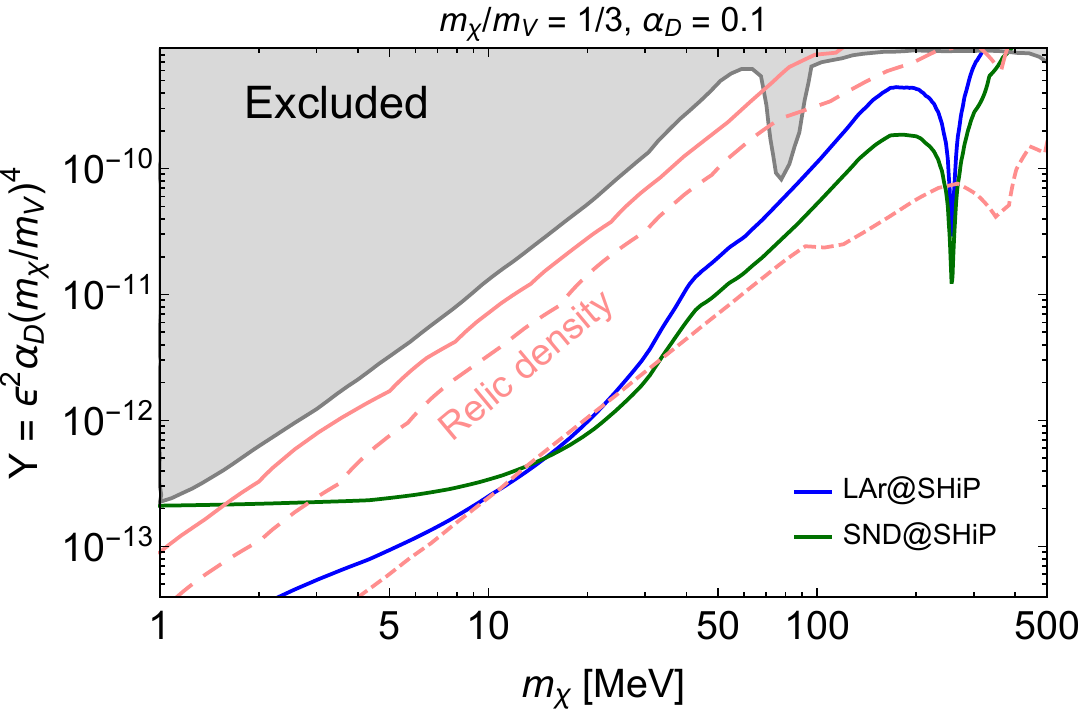}
    \caption{The sensitivity of the SND and LAr@SHiP detectors of the SHiP experiment to the elastic scattering of light dark matter $\chi$ coupled to dark photons off electrons. The full SHiP running time of 15 years (which is equivalent to the number of protons-on-target $N_{\text{PoT}} = 6\cdot 10^{20}$) is assumed in this figure and all the figures below. The light red lines correspond to the parameter space of the relic density of various minimal models of $\chi$ particles: complex scalar (solid), Majorana (long-dashed), and pseudo-Dirac (short-dashed)~\cite{SHiP:2020noy}.}
    \label{fig:sensitivity-LDM}
\end{figure}

Plugging in all relevant numbers in~\eqref{eq:rescaling-1} with the help of Table~\ref{tab:snd-vs-lar}, we get
\begin{equation}
    Y_{\text{lower,LAr}} \simeq 2 \sqrt{\frac{\sigma_{\text{SND}}}{\sigma_{\text{LAr}}}} Y_{\text{lower,SND}}
\end{equation}
The ratio of the cross-sections is $\ll 1$ for low dark photon masses $m_{V}\lesssim 50\text{ MeV}$, which is because the differential cross-section scales as $d\sigma/dE_{e} \propto 1/(m_{V}^{2}+2E_{e,\text{rec}}m_{e})^{2}\propto 1/2E_{e,\text{rec}}^{2}$ in this regime. For larger masses, the $d\sigma/dE_{e}$ has the asymptotic scaling $\propto 1/m_{V}^{2}$, and low recoil detection is no longer attractive -- the ratio of the cross-sections becomes $\gtrsim 1$.

The comparison of the sensitivities of SND@LHC and LAr@SHiP is shown in Fig.~\ref{fig:sensitivity-LDM}. We see the complementarity between the ability of the detectors to explore the parameter space of LDM, with LArTPC being able to probe better the domain of low masses and SND the range $m_{V}\gtrsim 50\text{ MeV}$.

\section{Millicharged particles}
\label{sec:mcp}
The interaction Lagrangian of millicharged particles (MCPs) 
\begin{equation}
    \mathcal{L} = \epsilon e \bar{\chi}\gamma^{\mu}\chi A_{\mu},
\end{equation}
where $\chi$ is the MCP, $A_{\mu}$ is the photon, and $\epsilon \ll 1$ is a small dimensionless parameter.

\begin{figure}[t!]
    \centering
    \includegraphics[width=0.65\textwidth]{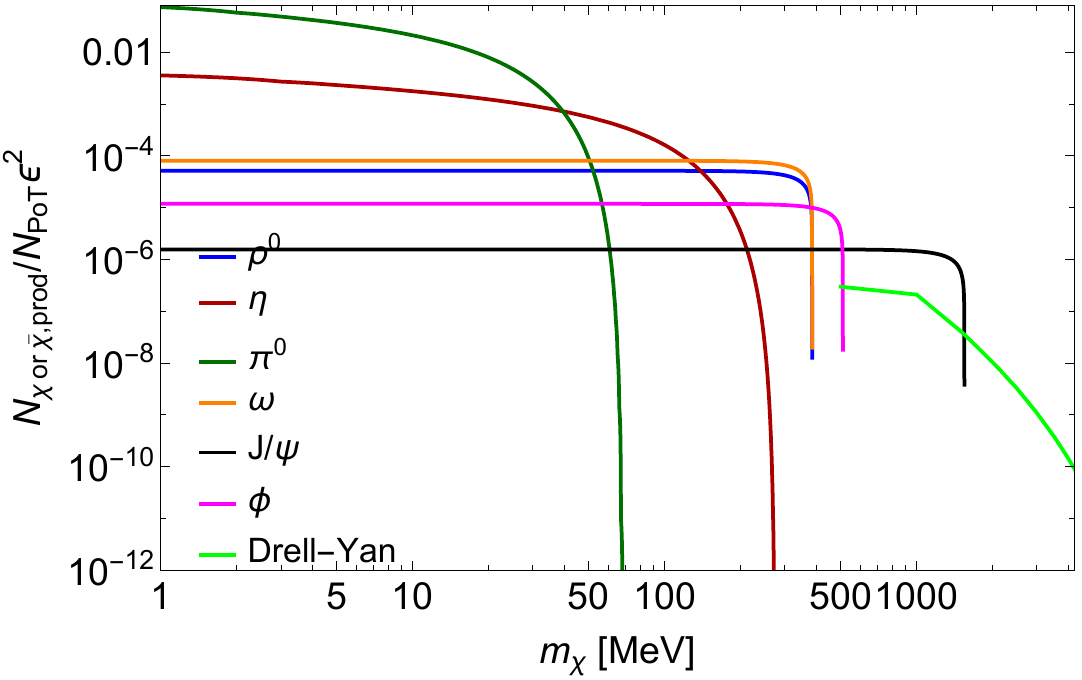}
    \caption{MCP mass dependence of the production probabilities for various mechanisms: 3-body decays of pseudoscalar mesons $\pi^{0}/\eta \to \gamma \chi\bar{\chi}$, 2-body decays of vector mesons $\rho^{0},\omega,\phi^{0},J/\psi \to \chi\bar{\chi}$, as well as the Drell-Yan process.}
    \label{fig:MCP-prod-probabilities}
\end{figure}

MCPs may be produced by 2- and 3-body decays of light mesons $\pi^{0},\eta,\eta',\rho^{0},\omega,J/\psi,\Upsilon$, as well as directly in proton-target collisions by the Drell-Yan process~\cite{Magill:2018tbb,Foroughi-Abari:2020qar}. The flux of the MCPs produced by these mechanisms has been calculated using \texttt{SensCalc}~\cite{Ovchynnikov:2023cry}. The production probabilities per proton-on-target (PoT), assuming the SHiP target, are shown in Fig.~\ref{fig:MCP-prod-probabilities}.

The possible signature is scatterings of MCPs inside the detector material. Unlike the case of FIPs interacting via a massive mediator for which the distribution of recoil electrons in the transferred momentum is flat below the mediator's mass, for MCPs, the electrons would likely have energies sharply peaked at small values. Therefore, searches for MCPs are a good objective for LAr detectors, where the possible energy threshold is well below 100 MeV. In Ref.~\cite{Harnik:2019zee}, it was proposed to search for MCPs via multiple soft interactions with electrons. In this case, the signature would be displaced hits with soft electrons along the trajectory of the MCPs pointing to the target. Such a signature has been used to constrain the parameter space of MCPs at the ArgoNeuT experiment~\cite{ArgoNeuT:2019ckq}, where the detectable electron energy recoil may be as small as $E_{e,\text{rec}}\simeq 1\text{ MeV}$. 

\subsection{SND@SHiP}
\label{sec:MCP-SND}
Similarly to LDM electron scattering searches, neutrino interactions with one visible electron at the primary vertex represent the main background to MCPs scattering signatures in the SND@SHiP environment. We adopt an analogous strategy to LDM studies, based on a two-step selection aimed at maximizing the significance of the MCPs scattering observation over the neutrino background, as defined in Sec.~\ref{sec:ldm-snd}. The optimized kinematic region of the scattered electron from MCPs is identified in $E_{e}\in[1,\,10]\,$GeV, $\theta_e\in[20,\,30]\,$mrad. A summary of residual backgrounds corresponding to $N_{p.o.t.}=6\times10^{20}$ is reported in Table~\ref{tab:nuyield_mcp}.
\begin{table}[t]
\centering
\begin{tabular}{lccccc}
\hline
& $\nu_{e}$ & $\bar{\nu}_{e}$ & $\nu_{\mu}$ & $\bar{\nu}_{\mu}$ & all\\ 
\hline
Elastic scattering on $e^{-}$ & 57 & 33 & 66 & 42 & 198 \\
Quasi~-~elastic scattering  & 81 & 93 &  &  & 174 \\
Resonant scattering  & - & 90 &  &  & 90 \\
Deep inelastic scattering  & - & - &  &  & -\\
\hline
Total & 138 & 216 & 66 & 42 & 462 \\
\hline
\end{tabular}
\caption{Neutrino background yield corresponding to the full SHiP running time of 15 years, which is equivalent to $6\times 10^{20}$ delivered PoT, assuming the selection criteria for MCP-electron elastic scattering searches.}
\label{tab:nuyield_mcp}
\end{table}

Unlike the case of LDM, the calculation of the SHiP sensitivity to MCPs has never been performed by the collaboration (see, however, Ref.~\cite{Magill:2018tbb}, which performed sensitivity studies for an old configuration without background studies). Therefore, we need to calculate the number of events from scratch. 

The number of events is given by
\begin{equation}
    N_{\text{ev}} = N_{\text{PoT}}\times \sum_{i}P_{\text{prod}}^{(i)}\times \int d\theta_{\chi} dE_{\chi} dE_{e} \ f_{\chi}^{(i)}(\theta_{\chi},E_{\chi})\epsilon_{\text{az}}(\theta_{\chi})\frac{dP_{\text{scatt}}}{dE_{e}}
    \label{eq:Nevents-LDM}
\end{equation}
Here, $N_{\text{PoT}}$ is the number of proton collisions. $P_{\text{prod}}^{(i)}$ is the probability to produce $\chi$ per proton collision:
\begin{equation}
    P_{\text{prod}} = 2\times \begin{cases} \frac{\sigma_{pp\to \chi\bar{\chi}}}{\sigma_{pp,\text{tot}}}, \quad \text{direct}\\ \chi_{X}\times \text{Br}(X\to \chi \bar{\chi} Y), \quad \text{secondary}
    \end{cases}
\end{equation}
where ``direct'' means the production directly in proton collisions (e.g., by the Drell-Yan process), and ``secondary'' means the production by decays of secondary particles $X$ with the amount per PoT being $\chi_{X}$. $f_{\chi}^{(i)}(\theta_{\chi},E_{\chi})$ is the FIP angle-energy distribution function normalized by 1. $z$ is the longitudinal displacement of the FIP from the production point. $\epsilon_{\text{az}} \equiv \Delta \phi(\theta,z)/2\pi$ is the geometric probability that the FIP's trajectory parametrized by $\theta,z$ lies inside the detector volume. Finally, $\frac{dP_{\text{scatt}}}{dE_{e}}$ is the differential scattering probability in the final electron energy $E_{e}$:
\begin{equation}
    \frac{dP_{\text{scatt}}}{dE_{e}} = n_{e,\text{LAr}} \Delta z_{\text{tg}}\times \frac{d\sigma}{dE_{e}}\times \epsilon_{\text{selection}},
    \label{eq:scattering-probability-ldm}
\end{equation}
where $z_{\text{tg}} = 1\text{ m}$, $n_{e} \approx 4.7\cdot 10^{30}\text{ m}^{-3}$ is the number density of electrons in the tungsten target,
\begin{equation}
    \frac{d\sigma}{dE_{e}} = \frac{8\pi \alpha_{\text{EM}}\epsilon^{2}m_{e}(E_{e}^{2}-2E_{e}E_{\chi}-3E_{e}m_{e}+2E_{\chi}^{2}+2E_{\chi}m_{e}+2m_{e}^{2}-m_{\chi}^{2})}{(E_{\chi}^{2}-m_{\chi}^{2})(2E_{e}m_{e}-2m_{e}^{2})^{2}}
    \label{eq:elastic-cross-section}
\end{equation}
is the differential cross-section, with $E_{e} = E_{e,\text{rec}}+m_{e}$, and $\epsilon_{\text{selection}}$ is the event selection cut:
\begin{equation}
    \epsilon_{\text{selection}}=h(
\theta_{\text{min}}<\theta_{e}(E_{\chi},E_{e})<\theta_{\text{max}})\times h(E_{\text{min}}<E_{e}<E_{\text{max}}),
\label{eq:selection-cut}
\end{equation}
with $h$ being the Heaviside step function.

The scaling of the number of events with the coupling $\epsilon$ is $N_{\text{ev}} \propto \epsilon^{2}\times \epsilon^{2} = \epsilon^{4}$, where the $\epsilon^{2}$ factors come from the production and scattering probabilities.

\subsection{LAr@SHiP}

For LAr, we will adopt the n-hit signature, where the MCP scatters several times, producing low-recoil electrons. The number of events has the form
\begin{equation}
    N_{\text{ev}} = N_{\text{PoT}}\times \sum_{i}P_{\text{prod}}^{(i)}\times \int d\theta_{\chi} dE_{\chi} \ f_{\chi}^{(i)}(\theta_{\chi},E_{\chi})\epsilon_{\text{az}}(\theta_{\chi})\langle P_{\text{scatt}}(E_{e,\text{thr}})\rangle,
    \label{eq:Nevents-LDM-lar}
\end{equation}
where
\begin{equation}
    \langle P_{\text{scatt}}\rangle = \frac{1}{n!}\left(\int \limits_{E_{\text{min}}} dE_{e}\ \frac{dP_{\text{scatt}}}{dE_{e}}\right)^{n}, \quad E_{\text{min}}=m_{e}+E_{\text{thr}}
\end{equation}
is the $n$-hit scattering probability (the expression~\eqref{eq:scattering-probability-ldm} with the parameters of the LAr detector), where by $E_{\text{thr}}$ we denote the minimal detectable recoil energy. The scaling of the number of events with $\epsilon$ and $E_{\text{thr}}$ is $N_{\text{ev}}\propto \epsilon^{2}\times (\epsilon^{2}/E_{\text{thr}})^{n}$.

\begin{figure}[t!]
    \centering
    \includegraphics[width=0.45\textwidth]{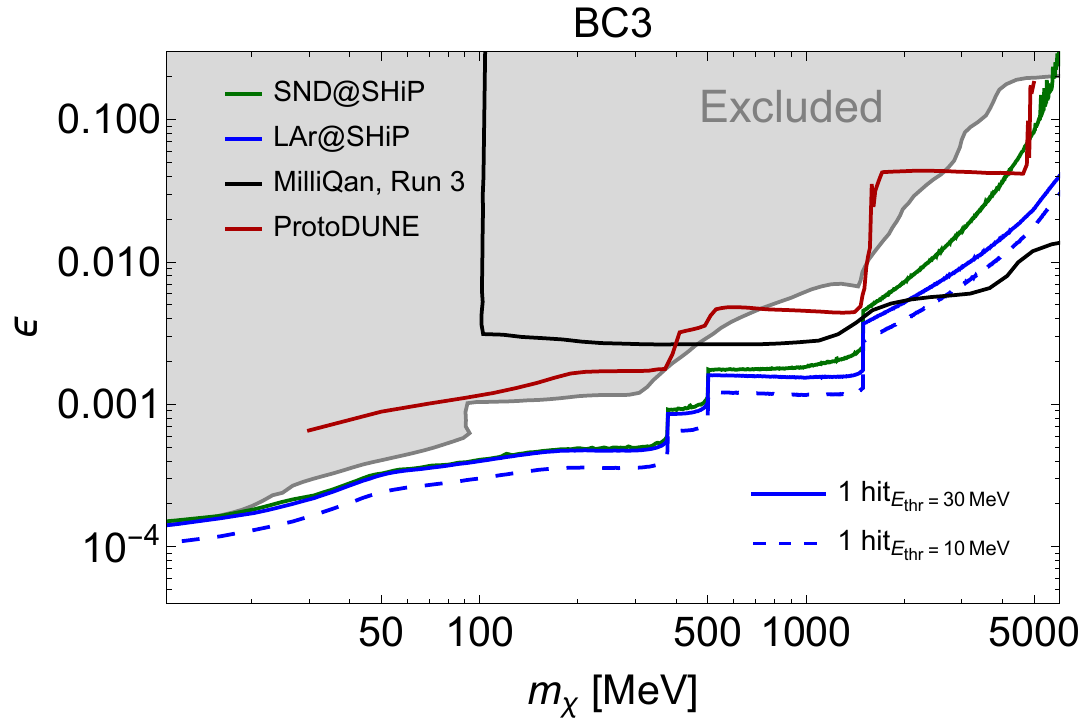}~\includegraphics[width=0.45\textwidth]{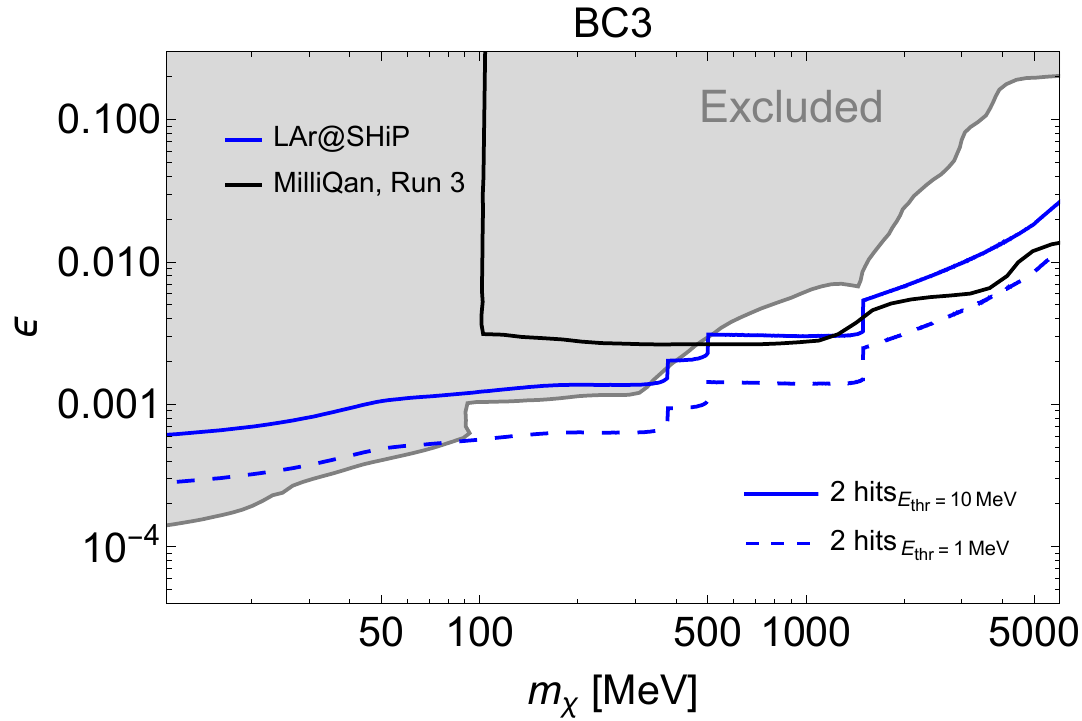}
    \caption{The sensitivity of SND and LAr detectors at SHiP to millicharged particles. \textit{Left panel}: the 90\% CL sensitivity curves of SND@SHiP, considering the background calculation described in the text (Sec.~\ref{sec:MCP-SND}) and LAr@SHiP, assuming the 1-hit signature with recoil thresholds $E_{\text{thr}} = 30\text{ MeV}$ and $10\text{ MeV}$ and the background from cosmic muons (see the present section). In the figure, we also show the ProtoDUNE sensitivity from~\cite{Coloma:2023adi} and the MilliQan sensitivity for LHC Run 3 statistics from~\cite{Antel:2023hkf}.  \textit{Right panel}: the 90\% CL sensitivity of LAr@SHiP, assuming 2-hit signature with thresholds $E_{\text{thr}} = 1,10\text{ MeV}$ and assuming the absence of backgrounds.}
    \label{fig:MCP-sensitivity-comparison}
\end{figure}

The 1-hit signature was adapted in the MCPs sensitivity study for proton fixed target experiments reported in~\cite{Kelly:2018brz}, and deemed to be sufficient as a signal when using scattered electron thresholds in the range of 10 MeV or more. We will assume two values of $E_{\text{thr}}$, namely, the nominal 30 MeV and --an optimistic-- 10 MeV value. To make an initial comparison with the other LArTPC proposals at SPS -- ProtoDUNE~\cite{Coloma:2023adi}, we consider the same background source and assumption as made in that paper -- cosmic background,  amounting to  $6.6\cdot 10^{6}$ for a  15-year running time of the SHiP experiment (Sec.~\ref{sec:selection-background}). The sensitivity of LAr to this signature is shown in Fig.~\ref{fig:MCP-sensitivity-comparison} (left panel), where we also include the sensitivity of MilliQan from~\cite{Antel:2023hkf} and SND@SHiP.

For $n= 2$, we consider two threshold values -- $E_{\text{thr}} = 1\text{ MeV}$ (similar to ArgoNeuT), and  $10\text{ MeV}$. We assume that the 2-hit signature is background-free. The expected sensitivity is shown in Fig.~\ref{fig:MCP-sensitivity-comparison} (right panel). 

We see that depending on the energy threshold, the 1-hit signature of the LAr option is as sensitive as the SND option. The sensitivities of these detectors are also above the MilliQan Run 3 sensitivity in the mass range $m_{\chi} \lesssim 1\text{ GeV}$, thanks to a much larger beam intensity, which leads to a larger MCP flux from light mesons such as $\rho,\omega,
\pi^{0}, J/\psi$. As for the 2-hit signature, it has a sensitivity competitive to the 1-hit, depending on the threshold choice, and simultaneously delivers the opportunity to identify the MCPs. Moreover, depending on the threshold, the sensitivity may be better than the sensitivity of MilliQan Run 3 even for large masses $m_{\chi} \sim 5\text{ GeV}$. 

\section{Inelastic light dark matter}
\label{sec:ldm-inelastic}
In this section, we consider the search for dark matter in a scenario where we have more than one matter particle in the dark sector. This will lead to an example of a double-bang topology. The model we are interested in is
\begin{equation}
    \mathcal{L}_{\text{int}} = i\sqrt{4\pi\alpha_{D}}\partial_{\mu}\chi_{2}V^{\mu}\chi_{1}^{*}+\text{h.c.}+V^{\mu}J_{\mu}
    \label{eq:inelastic-DM}
\end{equation}
Here, $\chi_{1,2}$ are scalar particles with $m_{\chi_{2}}>m_{\chi_{1}}$, and $\chi_{1}$ being stable. $V_{\mu}$ is a massive mediator coupled to a SM current $J_{\mu}$. $\chi_{1}$ may be a good light dark matter candidate. The relation of its abundance with the parameters in Eq.~\eqref{eq:inelastic-DM} depends on hidden assumptions such as the presence of the entropy dilution at some stage of the Universe's evolution. Therefore, we do not show the primordial abundance line in the final figures, assuming that in a very broad range of the parameter space $\chi_{1}$ may serve as DM or constitute its fraction. 

In principle, in addition to the off-diagonal interaction in~\eqref{eq:inelastic-DM}, there may also be diagonal interaction of the $\chi$ particles with the mediator. However, from the point of view of generic model building, it is possible to have a model where such types of interactions will be suppressed (see, e.g.,~\cite{Giudice:2017zke}). In this case, the direct DM detection experiments would not be able to probe the model, as the mass splitting between $\chi_{2},\chi_{1}$ would make the scattering of low-energy $\chi_{1}$ kinematically impossible (due to the absence of a $\chi_{1}-\chi_{1}$ coupling in the model).

Recently, Ref.~\cite{DeRoeck:2020ntj} proposed to search for the double bang events at the DUNE far detector with the boosted LDM produced in the atmosphere. In principle, the same signature may be used to search for accelerator-produced LDM. 

We will consider the interaction of the mediator with the baryon current:\footnote{The investigation for the dark photon mediator is left for future work.} 
\begin{equation}
    J_{\mu} = \frac{\sqrt{4\pi \alpha_{B}}}{3}\sum_{q}\bar{q}\gamma_{\mu}q,
    \label{eq:leptophobic}
\end{equation}
which corresponds to the case of the leptophobic mediator. Also, we will concentrate on the GeV scale for the mediator mass $m_{V}$. Several reasons dictate this choice. First, the missing energy search at the experiments with lepton beams like Belle II, BaBar, and NA64 cannot impose strong constraints on this model as there is no interaction with leptons. Second, LHC searches for the missing energy would be inefficient since they require a very large missing transverse energy/momentum, of the order of $100$ \text{ GeV}~\cite{Berlin:2018jbm}, which may be possible only if the mediator is that heavy. 

For the overview of the constraints on the leptophobic model and the phenomenology, see, e.g.,~\cite{Boyarsky:2021moj,Feng:2022inv} and references therein. The phenomenology is implemented in the code accompanying the paper. We assume the parameter space $m_{V}>m_{\chi_{1}}+m_{\chi_{2}}$, where, for definiteness, the second mass $m_{\chi_{2}}>m_{\chi_{1}}$. In this case, the production mechanism of the $\chi$ particles is 
\begin{equation}
    p p \to V + X, \quad V \to \chi_{1}+\chi_{2}
\end{equation} 
$\chi_{2}$ is unstable and quickly decays into $\chi_{1}$, and $\chi_{1}$ may reach LAr and scatter inside.

Let us now consider the double bang in detail. The first bang would consist of recoil hadrons from the $\chi_{1}$ scattering:
\begin{equation}
    \chi_{1}+p/n \to \chi_{2}+\text{recoil hadrons}
\end{equation}
For simplicity, we consider only the elastic scattering off protons. This way, the event rate estimate is conservative, as the deep-inelastic scatterings may constitute a huge fraction of events and even dominate the $\chi_{2}$ production. As for the second bang, the decay channels of $\chi_{2}$ are
\begin{equation}
\chi_{2} \to \chi_{1}+  \pi^{0}\gamma/\pi^{+}\pi^{-}\pi^{0},
\end{equation}
with the first one dominating in the mass range of interest. Therefore, the minimal mass splitting between $\chi_{2}$ and $\chi_{1}$ is $m_{\chi_{2}}-m_{\chi_{1}}>m_{\pi^{0}}$.

For the DB events, we will require the first bang energy threshold $E_{\text{rec}} > 50\text{ MeV}$ and the minimal displacement between the bangs of $L_{\text{min}} = 1\text{ cm}$. The latter corresponds to the expectations for the ability of the machine-learning algorithms to disentangle double bang from single bang.

\begin{figure}[t!]
    \centering
\includegraphics[width=0.65\textwidth]{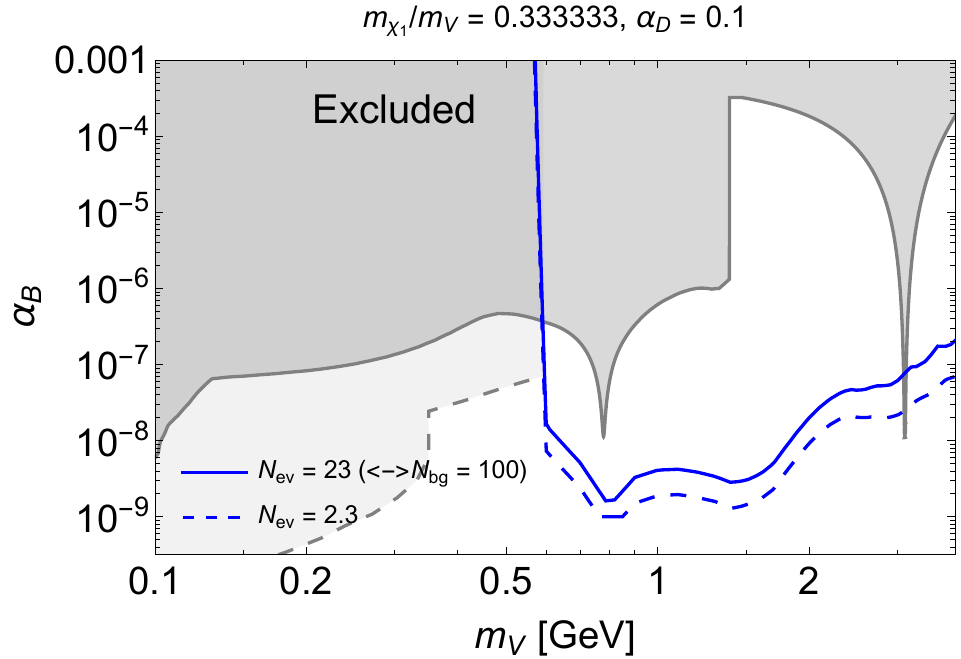}
    \caption{The potential of LAr@SHiP to explore the leptophobic portal via the double-bang signature, see text for details. The constraints are taken from~\cite{Boyarsky:2021moj}. The dashed line corresponds to the number of events $N_{\text{ev}} = 2.3$, while the solid to $N_{\text{ev}} = 23$, which is also equivalent to the 90\% CL sensitivity assuming 100 background events. The gray dashed line shows the UV-completion-dependent bounds from the anomaly-enhanced rate $B\to K +\text{inv}$ as computed in~\cite{Dror:2017nsg,Dror:2017ehi}. We show these constraints only below the threshold where $\chi_{2}$ may decay; above the threshold, the bounds must be recomputed.}
    \label{fig:double-bang-LDM}
\end{figure}

The number of events behaves as
\begin{equation}
    N_{\text{ev}} \approx N_{\chi_{1},\text{prod}}\times \int d\theta dE dE_{\text{rec}} f_{\chi_{1}}(\theta,E)\epsilon_{\text{az}}(\theta)\cdot n_{p,\text{LAr}}\cdot \frac{d\sigma_{\text{scatt}}}{dE_{\text{rec}}}\cdot \langle L\cdot P_{\text{decay},\chi_{2}}\rangle
    \label{eq:nev-inelastic-LDM}
\end{equation}
Here, $N_{\chi_{1},\text{prod}}\approx 2N_{V,\text{prod}}$ is the total number of the produced $\chi_{1}$ particles; $f_{\chi_{1}}$ is the angle-energy distribution of $\chi_{1}$; $\epsilon_{\text{az}}(\theta)$ is the azimuthal coverage of the LAr detector; $n_{p,\text{LAr}} \approx 3.9\cdot 10^{28}\text{ m}^{-3}$ is the number density of the protons in the LAr detector. $d\sigma_{\text{scatt}}/dE_{\text{rec}}$ is the differential scattering cross-section in the recoil energy $E_{\text{rec}} = E_{p}-m_{p}$, similar to the one for the elastic scattering~\eqref{eq:elastic-cross-section} but accounting for the mass difference between the incoming and outgoing $\chi$ particles and the elastic form-factor in the proton-leptophobic vertex, which we approximate to be the EM form-factor. Finally, 
\begin{equation}
    \langle L\cdot P_{\text{decay},\chi_{2}}\rangle \approx \int \limits_{0}^{L_{\text{det}}-L_{\text{min}}} dL \ \frac{L}{l_{\chi_{2},\text{decay}}}\times \exp\left[-(L_{\text{det}}-L)/l_{\chi_{2},\text{decay}}\right],
    \label{eq:decay-averaged}
\end{equation}
is the averaged decay probability accounting for the fact that the scattering probability increases with the length passed by the $\chi_{1}$ particle inside the detector, with $L_{\text{det}} = 9\text{ m}$. For simplicity, we have neglected the geometric limitations caused by the detector shape. $l_{\chi_{2},\text{decay}} = c\tau_{\chi_{2}}p_{\chi_{2}}/m_{\chi_{2}}$ is the decay length of $\chi_{2}$. The energy of the $\chi_{2}$ particle is related to the recoil energy as $E_{\chi_{2}} = E_{\chi_{1}}-E_{\text{rec}}$.

We will marginalize over the mass splitting $\Delta$. In practice, this means that we allow the decay length of $\chi_{2}$ to vary in wide ranges, controlled by $\Delta$. For $\Delta$ close to the kinematic threshold, the decay length of $\chi_{2}$ is much larger than $L_{\text{det}}$, and the $\chi_{2}$ particles mostly escape the detector acceptance. This means that the event observation probability is suppressed with the decay probability $L_{\text{det}}/c\tau_{\chi_{2}}\gamma_{\chi_{2}} \ll 1$. In the opposite case, when $\Delta$ is large, the $\chi_{2}$ particle decays instantly, and the event fails to meet the double bang displacement criterion. In this case, the event detection rate is exponentially suppressed with $\exp[-L_{\text{min}}/c\tau_{\chi_{2}}\gamma_{\chi_{2}}]$. In the intermediate regime, however, the decay probability within the detector range is $\mathcal{O}(1)$ and the displacement criterion is satisfied. In this case, the double bang signature may be as sensitive as the single scattering signature, with the benefit of a lower background for the DB.

The iso-event rate contours with the double-bang events rate assuming $\alpha_{D} = 0.1$ and the mass ratio $m_{\chi_{1}}/m_{V} = 1/3$ are shown in Fig.~\ref{fig:double-bang-LDM}. With LAr@SHiP, it is possible to go well beyond the parameter space excluded so far by past experiments.

\section{Dipole portal}
\label{sec:dipole}
In this section, we consider the sensitivity study for the HNLs coupled via the dipole portal.

The effective Lagrangian of the HNLs coupled to the SM via the dipole portal is
\begin{equation}
\mathcal{L} = d_{\alpha}\bar{N}\sigma^{\mu\nu}\nu_{\alpha,L}F_{\mu\nu},
\label{eq:dipole-portal}
\end{equation}
where $N$ is a HNL, $\nu_{\alpha}$ is an active neutrino, $F_{\mu\nu}$ is the EM strength tensor, and $d_{\alpha}$ is a dimensional coupling. The overview of phenomenology, constraints, and future searches may be found in Refs.~\cite{Magill:2018jla,Ovchynnikov:2022rqj,Ovchynnikov:2023wgg,Barducci:2023hzo}.

\begin{figure}[t!]
    \centering
    \includegraphics[width=0.7\textwidth]{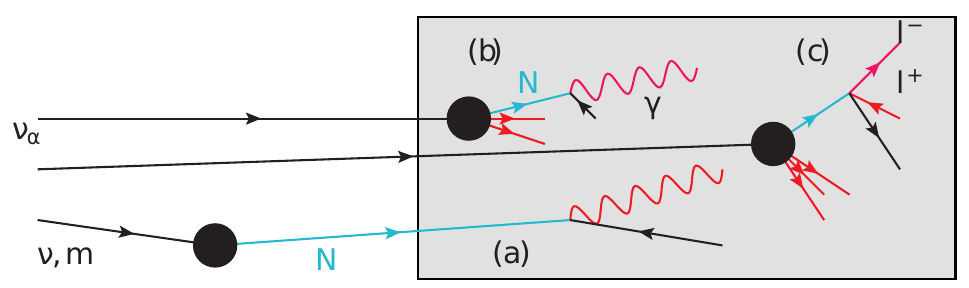}
    \caption{Various signatures with HNLs coupled via the dipole portal~\eqref{eq:dipole-portal}. The signature (a) corresponds to the HNLs produced outside the scattering detector, either by neutrino scattering or by decays of mesons and decaying inside (mostly into a photon and a neutrino). The signatures (b) and (c) correspond to the HNLs being produced by the neutrino up-scatterings inside the detector and then decaying into a photon or a pair of leptons within the detector. Depending on whether the recoil particles are visible, these events may be the double bang signature or a monophoton.}
    \label{fig:dipole-signature}
\end{figure}

Signatures with such HNLs depend on the place where they are produced -- inside or outside the LAr detector, see Fig.~\ref{fig:dipole-signature}. Let us briefly discuss the production channels (here we follow~\cite{Schwetz:2020xra}). The first mechanism is decays of short-lived mesons such as $\pi^{0},\eta,\eta', J/\psi$; they occur already inside the SHiP target. The second production channel is decays of long-lived mesons such as $\pi^{\pm},K^{\pm}$, which occur mainly inside the target or within the first few meters downstream, or muons, which may occur everywhere up to the LAr. Another important production channel is the up-scattering of the neutrinos; it may occur in the infrastructure upstream of the LAr detector (such as the hadron absorber, muon shield, and SND) or inside the LAr. 

We will include only the contributions from promptly decaying mesons and neutrino upscatterings inside the LAr, as the HNL flux from long-lived mesons and the neutrino-driven production significantly depends on the experimental setup, which is not finalized yet. Because of the same reason, we will concentrate on the case of the HNL coupling to $\tau$ neutrinos ($\alpha = \tau$ in Eq.~\eqref{eq:dipole-portal}). Indeed, unlike $\nu_{e/\mu}$, the $\nu_{\tau}$s are produced only promptly -- by the decays of $D_{s}\to \tau^{+} + \nu_{\tau}$ mesons and $\tau \to \nu_{\tau}+X$, and hence the estimates would be less setup-dependent. Both $\nu_{\tau}$ and $\bar{\nu}_{\tau}$ equally contribute to the HNL flux.

For the production outside the LAr, the only mechanism for the HNLs to manifest themselves is their decay. The main decay channel is $N\to \gamma + \nu$, which leads to a monophoton. The sub-dominant decay processes are $N\to l^{+}+l^{-}+\nu$, whose rate is suppressed by the extra photon vertex and the phase space of the 3-body decay but gets somewhat enhanced by the logarithmic factor $\ln(m_{N}/m_{l})$~\cite{Ovchynnikov:2022rqj}.   

For the production inside the LAr (via neutrino upscatterings), the signature depends on the scattering target --- electrons, nucleons, or nuclei. The detection signature can be either a double bang, with the recoil particle from the upscattering representing the first bang and the photon/di-lepton the second bang, or a single bang, if the recoil is too low to be detected. The latter situation is often the case for the scattering off nuclei, as the elastic nuclear form factor strongly suppresses large recoil thresholds. However, nuclear scattering dominates the production for undetectable recoils since the scattering probability gets enhanced by the factor of $2Z^{2}/A \sim 16$. 

To disentangle the sensitivity to these signatures, we will consider either the double bang signature with the lower threshold $E_{\text{rec}}>10\text{ MeV}$, or the monophoton and the di-lepton events, which we define as those with the upper bound on the recoil energy $E_{\text{rec}}<10\text{ MeV}$. Apart from the upscatterings with undetectable recoil inside the LAr, the latter events include decays of the HNLs produced outside the LAr. 

The number of events for the HNLs coupled to $\nu_{\alpha}$ and produced by the neutrino upscattering inside the detector has the form
\begin{equation}
    N_{\text{ev}} \approx N_{\nu_{\alpha}}\times \int d\theta dE dE_{\text{rec}} f_{\nu_{\alpha}}(\theta,E)\epsilon_{\text{az}}(\theta)\cdot \sum_{i = e,p,Z}n_{i,\text{LAr}}\cdot \frac{d\sigma_{i,\text{scatt}}}{dE_{\text{rec}}}\cdot \langle L\cdot P_{\text{decay},N}\rangle
    \label{eq:nev-dipole-upscattering}
\end{equation}
The expressions for the differential cross-sections $\frac{d\sigma^{i}_{\text{scatt}}}{dE_{\text{rec}}}$ (modulus the selection cut~\eqref{eq:selection-cut}) can be found in~\cite{Schwetz:2020xra}. The expression for $\langle L\cdot P_{\text{decay},N}\rangle$ is the same as in Eq.~\eqref{eq:decay-averaged} but with the replacement $\chi_{2}\to N$. For the double bang signature, we take $L_{\text{min}} = 1\text{ cm}$ in the expression for $\langle L\cdot P_{\text{decay},N}\rangle$ (similar to
the discussion in Sec.~\ref{sec:ldm-inelastic}) and $E_{\text{min}} = 50\text{ MeV}$ in~\eqref{eq:selection-cut}. For the monophoton and di-lepton signatures, we impose $E_{\text{min}} < 10\text{ MeV}$ and do not consider any cut on the displacement between the HNL production and decay vertex. For the distribution of $\nu_{\tau}$, we take the distribution of HNLs with zero mass and mixing with the $\tau$ flavor from \texttt{SensCalc}.
 
For the HNLs produced by decays of short-lived mesons, the number of events behaves as
\begin{equation}
    N_{\text{ev}} \approx \sum_{i = \pi^{0},\dots} N_{i}\times \text{Br}(i\to N)\times \int d\theta dz dE dE_{\text{rec}} f_{N}^{(i)}(\theta,E)\epsilon_{\text{az}}(\theta,z)\frac{\exp\left[-\frac{z}{l_{N,\text{decay}}}\right]}{l_{N,\text{decay}}}
    \label{eq:nev-dipole-meson}
\end{equation}
The branching ratios $\text{Br}(i\to N)$ may be found in~\cite{Ovchynnikov:2022rqj}.

\begin{figure}[t!]
    \centering
    \includegraphics[width=0.5\textwidth]{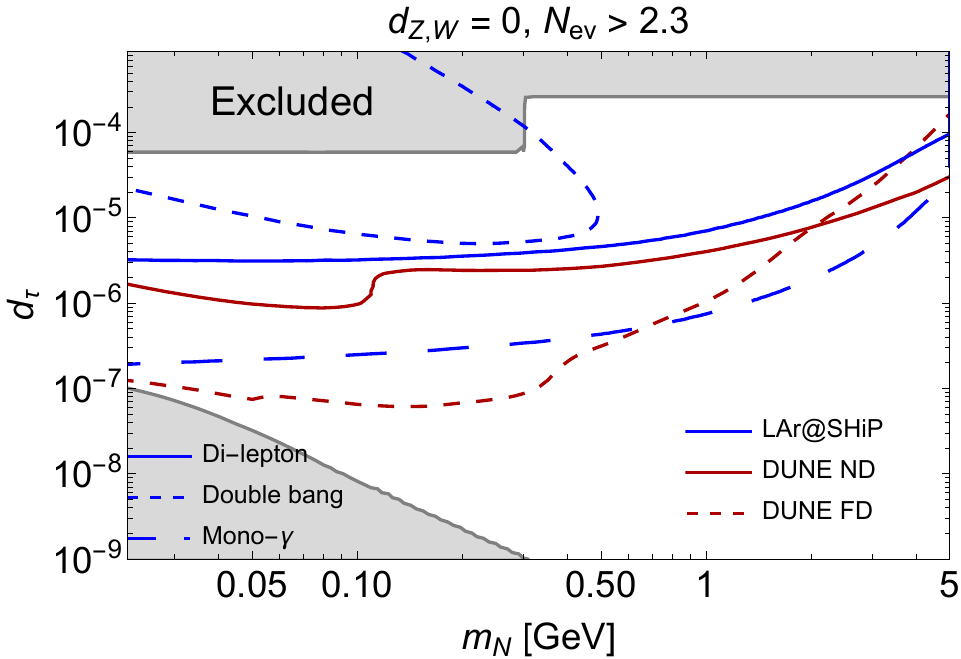}
    \caption{Iso-contours showing the sensitivity of LAr@SHiP to signatures with the HNLs coupled via dipole portal~\eqref{eq:dipole-portal} to $\nu_{\tau}$. The blue lines denote the parameter space to be probed with the di-lepton, mono-photon, and double bang signatures. For comparison, we also show the expected 10-year sensitivity of DUNE near and far detectors from~\cite{Ovchynnikov:2022rqj} to the monophoton signature, assuming the nominal horn configuration and $\simeq 100$ background events.}
    \label{fig:dipole-sensitivity}
\end{figure}

The sensitivity contours corresponding to these signatures are shown in Fig.~\ref{fig:dipole-sensitivity}. The parameter space that can be covered with a double bang signature is limited; the main reason is the smallness of the HNL decay length as well as the preference for tiny recoil events in the case of the HNL dominant production channels - scatterings from nuclei. The dominant signature is anticipated to be the monophoton one, but detailed background studies in the future are needed to confirm this.

In the same figure, we include the sensitivity of DUNE as calculated in~\cite{Ovchynnikov:2022rqj}. The DUNE setup used to obtain the sensitivity is as follows. A 10-year running time with the total number of protons-on-target $N_{\text{PoT}} = 1.1\cdot 10^{22}$ and the nominal focusing horn configuration were assumed. For the near detector, a LAr setup with dimensions $3\times 6 \times 2 \text{ m}^{3}$ was used. All four modules of the far detector have been included when obtaining the event rate.

\section{Conclusions}
\label{sec:conclusions}

The SPS at CERN delivers an excellent opportunity for the search for long-lived particles (LLPs), combining the huge intensity of the incoming proton beam with a relatively large proton energy. This feature will be exploited by the proposals of the experiments to be installed at the ECN3 facility. 

In this paper, we have explored the potential of SPS to search for various signatures with LLPs by considering adding a liquid argon detector and, as a
concrete example, have it located behind the SHiP spectrometer (Sec.~\ref{sec:lar}). The LAr setup provides excellent capabilities in timing and vertex resolution, as well as charged track and energy measurements, which may be used to study the signatures that may either complement the other SHiP detectors -- the SND and hidden sector decay spectrometer -- for the models already explored or to open the unique opportunity to probe entirely different models, see Sec.~\ref{sec:LAr-opportunities}. Examples include the visualization of the event, which is especially important for LLP scatterings or many-body decays (such as hadronic decays), multi-hit signatures, or low-recoil scatterings. These signatures may be accompanied by low or even negligible background, although a detailed study may be finalized only after finalizing the optimization of the experimental
setup (such as the magnetic shield design), as well at the proposed LArTPC, Sec.~\ref{sec:selection-background}.

We have considered a few case studies with various LLPs -- light dark matter coupled to dark photons, inelastic light dark matter (LDM) coupled to the leptophobic mediator, millicharged particles (MCPs), and heavy neutral leptons coupled via the dipole portal.

For the case of the LDM coupled to dark photons (Sec.~\ref{sec:ldm}) and MCPs (Sec.~\ref{sec:mcp}), we have studied the sensitivity of the updated SND@SHiP setup and LAr@SHiP. For the former, we have found that the detection of low-recoil events at LAr may significantly extend the sensitivity of SND@SHiP to the domain of small dark photon masses, $m_{V}\lesssim 50\text{ MeV}$, see Fig.~\ref{fig:sensitivity-LDM}. For the millicharged particles, Sec.~\ref{sec:mcp}, one can use a single-hit or multi-hit signature -- a few MCP scatterings with low-recoil electrons, with the trajectory pointing to the target. For reaching the same sensitivity with multi-hit signatures, the threshold for each hit needs to be much lower, though, which will need to be demonstrated experimentally. The sensitivity is comparable to the single-hit sensitivity of SND@SHiP but allows distinguishing MCPs from other hypothetical LLPs. It may also go beyond the parameter space to be covered by MilliQan in Run 3 (Fig.~\ref{fig:MCP-sensitivity-comparison}).

For the models of inelastic LDM (Sec.~\ref{sec:ldm-inelastic}) and HNLs (Sec.~\ref{sec:dipole}), the suitable signature may be a double bang, with the first bang being the low-recoil scattering producing the unstable particle that then decays with a large energy release after passing a macroscopic distance (see Fig.~\ref{fig:double-bang}). Such signatures would allow not only the identification of the model but also -- in the case of many observed events -- reconstruct the decay length of the decaying LLP. For HNLs, the parameter space to be covered with the double bang events is significantly limited compared to the more ``standard'' signatures with the other possible signatures -- isolated di-lepton and monophoton events (Fig.~\ref{fig:dipole-signature}), see Fig.~\ref{fig:dipole-sensitivity}. For the inelastic LDM, the situation is different (Fig.~\ref{fig:double-bang-LDM}): if marginalizing over the mass splitting between the dark matter particle and its heavier unstable counterpart, the event rate for the double bang signature may be as high as for the single-event signature, with the benefit of a lower background.

To summarize, the LAr option would nicely complement the existing ECN3 experimental proposals and significantly push its capabilities in the range of LLP identification.

\section*{Acknowledgements}
The authors would like to thank G. De Lellis, A. Di Crescenzo, and R. Jacobsson for fruitful discussions. MO has received support from the European Union’s Horizon 2020 research and innovation program under the Marie Sklodowska-Curie grant agreement No. 860881-HIDDeN. The support from the Swiss National Science Foundation (SNSF) under Contract No. $200020\_204238$ is acknowledged.

\bibliographystyle{JHEP}
\bibliography{bib}

\newpage

\appendix

\newpage 
\appendix

\section{Rescaling the SHiP sensitivity to LDM}
\label{app:rescaling}

The scaling of the number of events $N_{\text{ev}}$ with couplings has the form
    \begin{equation}
     N_{\text{ev}} \propto N_{\text{prod,tot}}\times \epsilon_{\text{geom}}\times P_{\text{scatt}} \propto \epsilon^{2}\times\epsilon_{\text{geom}}\times Z_{\text{tg}}\cdot n_{\text{tg}}\cdot \Delta z_{\text{tg}}\epsilon^{2}\langle \sigma_{\text{scatt}} \rangle
    \label{eq:scaling-1}
    \end{equation}
Here, $\epsilon_{\text{geom}}$ is the fraction of $\chi$ flying to the detector, $\langle\sigma_{\text{scatt}}\rangle = \epsilon^{2}\alpha_{D}f(m_{\chi},m_{V})$ is the cross-section averaged over $\chi$ angles and energies.  $Z_{\text{tg}}$ is the target's charge (accounting for the number of electrons per nucleus), and $n_{\text{tg}}$ is the atomic number density. Finally, $\Delta z_{\text{tg}}$ is the length of the target inside the detector. 

Since SHiP is located in the forward direction, the solid angle $\chi$ distribution is flat. As a result, for geometric acceptance, we may use 
\begin{equation}
\epsilon_{\text{geom}}\propto \Omega_{\text{det}} = \frac{S_{\text{det},\perp}}{z_{\text{to det}}^{2}}
\label{eq:scaling-2}
\end{equation}
Also, it is reasonable to assume that the $\chi$ energy spectrum does not depend on the SHiP configuration. Hence, $f(m_{\chi},m_{V})$ in the cross-section is setup-independent.

Next, let us express $\Delta z_{\text{tg}}$ in terms of parameters of the experiment -- the detector length $l_{\text{det}}$, its total mass $m_{\text{det}}$, the atomic number of the target $A_{\text{tg}}$, and the total volume $V_{\text{det}}$:
\begin{equation}
\Delta z_{\text{tg}} \approx \Delta z_{\text{det}}\cdot \frac{m_{\text{det}}}{\rho_{\text{tg}}V_{\text{det}}} \propto \Delta z_{\text{det}}\cdot \frac{m_{\text{det}}}{A_{\text{tg}}n_{\text{tg}}V_{\text{det}}}
\label{eq:scaling-3}
\end{equation}
Combining Eqs.~\eqref{eq:scaling-1}-\eqref{eq:scaling-3}, we get
\begin{equation}
    N_{\text{ev}} \propto \frac{Z_{\text{tg}} m_{\text{det}}}{A_{\text{tg}}z_{\text{to det}}^{2}}
\label{eq:Nev-scaling-1}
\end{equation}
Finally, let us derive the scaling of the upper bound of the sensitivity. The number of events scales with the couplings as 
\begin{equation}
N_{\text{ev}}\propto \alpha_{D}\epsilon^{4} \propto Y^{2}/\alpha_{D}
\end{equation}
Requiring that $N_{\text{ev}}>2.3\sqrt{N_{\text{bg}}}$ with $N_{\text{bg}}$ being the background number, for the lower bound of the sensitivity we get
\begin{equation}
    Y_{\text{lower}} \propto \left(\frac{A_{\text{tg}}\sqrt{N_{\text{bg}}}z^{2}_{\text{to det}}}{Z_{\text{tg}}m_{\text{det}}}\right)^{\frac{1}{2}}
\end{equation}

\end{document}